\documentstyle[aps,epsf,prb,multicol]{revtex}
\renewcommand{\vec}[1]{\mbox{\boldmath$#1$}}
\newcommand{\vecscr}[1]{\mbox{\boldmath$\scriptstyle #1$}}
\newcommand{\scr}{\scriptsize}

\renewcommand{\arg}{\mbox{Arg}\,}

\title{Quasiparticle Spectrum of $d$-wave Superconductors \\ in the Mixed
State}
\author{Luca\ Marinelli, B.\ I.\ Halperin \\ \small \emph{Physics
Department, Harvard University, Cambridge, MA 02138} \\ \normalsize S.\
H.\ Simon \\ \small \emph{Lucent Technologies Bell Labs, Murray Hill,
NJ 07974}}
\date{January 27, 2000}

\begin{document}
\maketitle

\begin{abstract}
The quasiparticle spectrum of a two-dimensional $d$-wave superconductor in the 
mixed
state, $H_{c1} \ll H \ll H_{c2}$, is studied both analytically and
numerically using the linearized Bogoliubov--de Gennes
equation.  We consider various values of the ``anisotropy ratio"
$v_F/v_\Delta$ for the quasiparticle velocities at the Dirac points,
and we examine the implications of symmetry. For a Bravais lattice of 
vortices, we find there is always an isolated energy-zero
(Dirac point) at the center of the Brillouin zone, but for a
non-Bravais lattice  
with two vortices per unit cell there is generally an energy gap.
In both of these cases, the density of 
states should vanish at zero energy, in contrast with the semiclassical 
prediction of a constant density of states, though the latter may hold
down to very low energies for large anisotropy ratios. This result is
closely related to the particle-hole symmetry of the band structures
in lattices with two vortices per unit cell. More
complicated non-Bravais vortex lattice configurations with at least
four vortices per unit cell can break the particle-hole symmetry of
the linearized energy spectrum and lead to a finite density of states
at zero energy. 
\end{abstract} 

\begin{multicols}{2}

\section{Introduction}
In the past few years several key experiments have demonstrated that
the order parameter in high-temperature superconductors has $d$-wave
symmetry rather than the conventional $s-$wave symmetry found in
low-temperature superconductors \cite{vanharlingen95}. This fact has
strong implications for the low-temperature thermodynamics of these
systems, because the $d$-wave symmetry implies the existence of points
on the Fermi surface where the gap function vanishes. At these points
the bulk quasiparticle spectrum will be gapless and these states will
be occupied even at very low temperature. 

High-$T_c$ superconductors
are extreme Type-II superconconductors and in a magnetic field larger
than $H_{c1}$ develop a vortex lattice. The geometry of this vortex
lattice has been investigated via small angle neutron scattering
\cite{keimer94}, scanning tunneling microscopy
\cite{maggioaprile95,renner98} and magnetic decoration \cite{kim99} in
YBa$_2$Cu$_3$O$_7$ (YBCO) and Bi$_2$Sr$_2$CaCu$_2$O$_8$ (Bi2212). 
In YBCO twinned single 
crystals the vortex lattice in a magnetic field of 6 T parallel to the
$c$ axis looks like a
skewed square lattice with an angle between primitive vectors of about
$77^\circ$ \cite{maggioaprile95}. Low-field magnetic decoration
studies of Bi2212 between 70 G and 120 G parallel to the $c$ axis
\cite{kim99} find a vortex lattice very close to a hexagonal one.

We have studied the quasiparticle spectrum of a $d$-wave
superconductor in the vortex lattice within the framework of the
Bogoliubov--de Gennes equation. The main question we have addressed is
whether the spectrum becomes gapped in the presence of a magnetic
field $H_{c1} \ll H \ll H_{c2}$ and more generally what the energy spectrum 
looks like. This
question was previously approached via numerical simulation 
of a tight-binding model \cite{wang95,yasui99} and semiclassical analysis
\cite{volovik93}. Gor'kov and Schrieffer \cite{gorkov98} and, in a more
recent preprint using different arguments, Anderson \cite{anderson98}
predicted that the quasiparticle spectrum of a $d$-wave superconductor
in a magnetic field $H \ll H_{c2}$ is characterized by broadened Landau
levels with energy levels
\begin{equation}
E_n = \pm \hbar \omega_H \sqrt{n}, \ \ n=0,1,\ldots
\label{eq:landaulevel}
\end{equation}
where $\omega_H = \sqrt{2 \omega_c \Delta_0/\hbar}$. Here $\omega_c =
|e H|/mc$ is the cyclotron frequency and $\Delta_0$ is the maximum
superconducting gap.
A key assumption of Gor'kov and Schrieffer and of Anderson, however, was to
neglect the spatially dependent superfluid velocity which has been
shown to strongly mix Landau levels by Mel'nikov \cite{melnikov99}.  

Recently a preprint by Franz and Te\v{s}anovi\'{c}
\cite{franz00} has given new insight into the problem. They introduced
a gauge transformation that takes into account the supercurrent
distribution and the magnetic field on an equal footing. In this way
they map the original problem onto that of diagonalizing a Dirac
Hamiltonian in an effective periodic vector and scalar potential with
vanishing magnetic flux in the unit cell. Employing the Franz-Te\v{s}anovi\'{c}
transformation, we were able to tackle the problem of understanding the band
structure of this system, via both analytic and numerical methods.

Our analysis will be limited to the spectrum at low energies, in the case where
the magnetic field is very small compared to $H_{c2}$.  In this case the
distance between the vortices is large compared to their diameter, and we can
ignore contributions to the spectrum from inside the vortex cores.  
The quasiparticle states
of interest to us are then constructed from excitations close to the
nodes of the
energy gap of the zero-field spectrum, and we can ignore mixing
between different
nodes.  Thus we analyze the effects of the magnetic field and vortex
lattice in a model where the zero field spectrum consists of four independent 
anisotropic Dirac cones. (We assume that the magnetic field itself is uniform in
the sample, which is appropriate for a bulk superconductor provided that 
$H \gg H_{c1}$, but extends to even smaller fields for a very thin sample.)

A very important numerical parameter in the problem is 
the anisotropy ratio $\alpha_D =
v_F/v_\Delta$. Here $v_F$ is the Fermi velocity, while
$v_\Delta = \Delta_0/p_F$ is the quasiparticle velocity parallel to the Fermi
surface, so the ratio $\alpha_D$
measures the anisotropy of the quasiparticle velocities at each Dirac
point, in zero field. From angle-resolved photoemission spectroscopy
and thermal conductivity measurements, the value of $\alpha_D$ for
high-$T_c$ superconductors turns out to be about 14 for YBCO
\cite{chiao99} and 20 for Bi2212 \cite{chiao99,mesot99}.  
Conceptually, however,
it is useful to consider the entire range of possible values for $\alpha_D$,
including the ``isotropic case" $\alpha_D = 1$.  This is particularly useful
because we find that certain features of the spectrum, such as energy gaps can
become extremely small for large values of the anisotropy, and therefore become
difficult to resolve numerically.

Since each vortex in a superconductor carries 
only carry half of the normal flux quantum $\Phi_0 = hc/|e|$, it is
necessary to
choose a unit cell with an even number of vortices, so that the electron
wavefunctions are single valued.
Thus, if the vortices sit on a Bravais lattice with one vortex per unit
cell, it is necessary to use a double unit cell, containing two vortices, in
order to carry out the analysis.  On the other hand, if the vortices sit on a
non-Bravais lattice, with two vortices per unit cell, one can use directly 
the unit cell of the vortex lattice.

Some key features of the quasiparticle band structure may be noted by looking 
at
Figs.~\ref{figdelta1}-\ref{figdelta4}
below, which give results of our numerical diagonalization (discussed in
Section~\ref{sect:numband}) for a  square Bravais lattice, 
rotated by $45^\circ$ from the
quasiparticle anisotropy axis,  with relatively small anisotropy
ratio $1 \le \alpha_D \le 4$. One striking feature 
is the presence of band crossings both at zero and finite energy,
regardless of the value of the anisotropy ratio $\alpha_D$. 
At least for relatively small anisotropy, the level crossings seem to be
limited to the $\Gamma$ and M point, as defined in Fig.~\ref{figAB}.
These band crossings are particularly interesting because in the absence of
some special symmetries,  we would expect their probability to vanish for a
two-dimensional Hamiltonian with broken time-reversal symmetry, as
is discussed in more detail in Section~\ref{sect:pertgen}. 

We have studied the role played by various symmetries of the Hamiltonian
both to simplify our numerical computations and to try to understand
why band crossings are allowed at some isolated points in the magnetic
Brillouin zone. In particular, we focused on zero-energy states, as
their existence changes qualitatively the thermodynamic functions of
the system, at very low temperatures.  For the lattices described above, we 
find exact particle-hole symmetry, both numerically and
analytically, at each point of the magnetic Brillouin zone, as can be
seen from the plotted band structures and is further discussed in
Section~\ref{sect:symm}. Doing perturbation theory calculations
described in Section~\ref{sect:pertgamma}, we also found that a crucial
role is played by the Bravais nature of the vortex lattice. This
symmetry, together with particle-hole symmetry is directly responsible
for the spectrum staying gapless in the presence of a magnetic
field. To prove this, we considered, both in perturbation theory and
by numerical diagonalization, what happens if we deform the vortex
lattice so that the distance between the $A$ and $B$ sublattices
$2 R_0$ defined in Fig.~\ref{figAB} is not 1/2 of the distance between two
flux lines belonging to the same sublattice, measured along the
diagonal. We found that particle-hole symmetry still exists at each
point, separately, of the Brillouin zone but that, generally, gaps
open up both at the $\Gamma$ and M
points, as can be seen in Fig.~\ref{figp}. This result will be
discussed in more detail in Section~\ref{sect:epsilon}.

Another important question is whether there are any further
zero-energy modes in addition to the ones at the $\Gamma$ point. This
is a very delicate 
question to address numerically because it is hard to distinguish
small gaps from real zero-energy eigenvalues, both because of finite
numerical accuracy and, more importantly, because of finite grid-size
effects. The latter can be particularly troublesome when dealing with
lattice fermions as will be discussed at the beginning of
Section~\ref{sect:numband}. In numerical calculations, 
as noted by Franz and Te\v{s}anovi\'{c},
it appears that for anisotropies of the order of 15, there are two lines
in the Brillouin zone where the quasiparticle energy vanishes.  Based on our
symmetry analysis, however, we conclude that there actually remains a
very small energy gap all along these lines, both for the case of a
Bravais lattice and for a non-Bravais lattice with two vortices per
unit cell. On the other hand, we find that for ``rectangular'' vortex lattices,
isolated energy zeroes are allowed along the $\Gamma$X symmetry line
(or -X$\Gamma$, by inversion symmetry).

Different conclusions are reached if one allows for more
complicated lattice structures, for example considering four vortices
per unit cell. In this case it is possible to have superfluid velocity
distributions without a center of inversion symmetry leading to 
non particle-hole symmetric energy spectra as shown in
Fig.~\ref{figfournosym}. For large enough
anisotropy, there is nothing that prevents lines of energy zeroes from
appearing, and the density of states can become finite at zero energy.  

The scaling laws for thermodynamic functions computed by Volovik 
\cite{volovik97} within the semiclassical theory, and by Simon and Lee 
\cite{simon97} in a framework closer to our approach, have been tested
experimentally (see for example \cite{revaz98}). It is of interest to 
determine the relation between the semiclassical approximation and the full
quantum mechanical spectrum. We have extracted a density of states from
the band structures we computed for square vortex lattices and,
contrary to the semiclassical
prediction of a constant density of states at zero energy, we find a
linearly vanishing density of states at low-energy for anisotropy
ratio $\alpha_D \leq 4$, as shown in
Figs.~\ref{figdelta1}-\ref{figdelta4}. Because of the previous
discussion on the non-existence of lines of energy zeroes, we are led to
believe that this result holds for any value of the anisotropy ratio
as long as the vortex lattice has one or two vortices per unit cell.
However the semiclassical approximation may be valid down to extremely
low energies, when $\alpha_D$ is large.
 
In Section~\ref{sect:dos} we study the
crossover between the semiclassical and quantum mechanical
regions. There are two relevant energy scales which we call $E_1$ and
$E_2$. For $E>E_1$ the density of states is qualitatively identical to
the bulk density of states in the absence of a magnetic field,
although there are still noticeable features induced by van Hove
singularities of the band structure in the vortex lattice. Below
$E_1$, the presence of a 
magnetic field can be accounted for within a semiclassical
approximation, all the way down to an energy scale $E_2$ where a full
quantum mechanical calculation becomes necessary. 

We compare our numerically determined energy scales $E_1$ and $E_2$
with the crossover scales introduced by
Kopnin and Volovik  \cite{kopnin96,volovik97} (see also
\cite{volovik97b}) $E_1 \approx \hbar v_F/d$ and  
$E_2^{KV} \approx \hbar v_\Delta/d$, where $d$ is the average distance
between vortices. While we agree with their expression for $E_1$,
we notice that for large anisotropy our numerical analysis 
indicates that, at least for the geometries considered here, $E_2$
should go to zero much faster than $1/\alpha_D$. 
(The precise functional dependence of $E_2$ on the anisotropy ratio
$\alpha_D$ will be the object of a paper currently in preparation.)

The two energies $E_1$ and $E_2^{KV}$ can also be written as $E_1 \approx
T_c \sqrt{H/H_{c2}}$ and 
$E_2^{KV} \approx E_1 (T_c /E_F) = (T_c^2/E_F) \sqrt{H/H_{c2}}$. 
Experimentally, 
taking $v_F = 2.5 \times 10^7 \mbox{cm/s}$ \cite{mesot99} we find that 
$E_1/(k_B \sqrt{H}) = 30\: \mathrm{K/T}^{1/2}$ and $E_2^{KV}/(k_B \sqrt{H}) = 
2 \:\mathrm{K/T}^{1/2}$ for both YBCO and Bi2212, to a good approximation. Recent
specific heat \cite{revaz98} and especially low temperature thermal
conductivity \cite{chiao99_2} experiments have been performed in the
regime $E < E_2^{KV}$ and still show good agreement with the semiclassical
predictions, which also seems to suggest that the right crossover
scale between the quantum mechanical and semiclassical regime is much
smaller than $E_2^{KV}$, for large anisotropy ratios. Note that the
Landau level energy scale (\ref{eq:landaulevel}) of Gor'kov and
Schrieffer and of Anderson is approximately the geometric mean of
$E_1$ and $E^{KV}_2$.

For $E > E_1$, the density of states is linear in energy
with superimposed sharp peaks (logarithmic van Hove singularities) and
the slope is the same as the one in zero magnetic field. 
When $E_2< E < E_1$, we are in the regime where
the semiclassical theory predicts a constant density of states. This
region shrinks as one lowers the anisotropy ratio $\alpha_D$, 
and disappears entirely in the isotropic limit $\alpha_D =1$
as is apparent by looking at Fig.~\ref{figdelta1}. In the very
low-energy limit $E< E_2$, the semiclassical theory breaks down and
the density of states has to be computed through the quantum
mechanical spectrum. We find that for $E \ll E_2$ the density of
states is linear and vanishes at zero energy, for the Bravais latitce. 

To summarize the structure of the paper, the model Hamiltonian is
discussed in Sections \ref{sect:BdG}-\ref{sect:delta}. An analysis of
particle-hole symmetry follows in Section \ref{sect:symm}. In Section
\ref{sect:numband} we describe the numerical methods and results of the
band structure calculation for a Bravais vortex lattice, while in Sections
\ref{sect:pertgamma}-\ref{sect:pertgen} we study the low-energy states in
perturbation theory. More general vortex lattices with two and four
vortices per unit cell are considered in Section \ref{sect:epsilon}. 
In Section \ref{sect:dos}, we focus on the density of states and compare
it to the semiclassical predictions.  Conclusions follow in 
Section~\ref{sect:concl}.

\section{Linearized Bogoliubov--de Gennes equation}
\label{sect:BdG}
In a spatially inhomogeneous system, the standard 
approach to the
description of the quasiparticle spectrum is provided by the
Bogoliubov--de Gennes equation \cite{degennes89}. For an arbitrary gap
operator (i.e. not necessarily $s-$wave), it reads ${\cal H}_{\mbox{\scr
BdG}} \psi = E
\psi$ where $\psi = (u,v)^T$ is a Nambu 2-spinor whose components are
the particlelike and holelike part of the quasiparticle wave function,
respectively. The Bogoliubov--de Gennes operator ${\cal H}_{\mbox{\scr
BdG}}$ we will consider is \cite{simon97}
\begin{equation}
\label{bdgeq}
{\cal H}_{\mbox{\scr BdG}} = \left(
\begin{array}{cc}
\displaystyle \frac{(\vec{p}-\frac{e}{c}\vec{A})^2}{2 m} - E_F &
\hat{\Delta} \\ 
\hat{\Delta}^* & \displaystyle
-\frac{(\vec{p}+\frac{e}{c}\vec{A})^2}{2 m} +E_F  
\end{array}
\right),
\end{equation}
where $E_F$ is the Fermi energy and $m$ is the electron effective
mass. Notice that we are neglecting the
self-consistent interaction potential \cite{degennes89} (analogous to
the Hartree--Fock potential in the normal phase) and any disorder
potential. The gap operator
acts on components of the wave function as $\hat{\Delta} g(\vec{r}) =
\int d\vec{r}^\prime\, \Delta_d(\vec{r},\vec{r}^\prime)
g(\vec{r}^\prime)$. For a $d_{xy}$ superconductor, the gap operator
can be expressed
as $\hat{\Delta} = \frac{1}{p_F^2} \{p_x,\{p_y,\Delta(\vec{r})\}\}$
where $\Delta(\vec{r}) = \Delta_0 e^{i \phi(\vecscr{r})}$ and the
brackets represent symmetrization $\{a,b\} = \frac{1}{2}
(ab+ba)$. We choose this orientation instead of the more conventional
$d_{x^2-y^2}$ purely for notational simplicity; results do not
depend on this choice. 

In the absence of a magnetic field, there are four points on the
Fermi surface at $\vec{p}=(\pm p_F, 0)$ and $\vec{p} = (0, \pm p_F)$
where the gap vanishes. If we are interested in the low-energy
properties of the quasiparticle excitation spectrum, we can linearize
the Bogoliubov--de Gennes equation around one of these points. This procedure
is justified because we are considering magnetic fields $H \ll H_{c2}$
and so the inverse magnetic length is much smaller than
$k_F$. We can choose to linearize around $\vec{p} = (0,p_F)$ writing
the wave function
$\psi = e^{i k_F y} \tilde{\psi}$. The Bogoliubov--de Gennes
equation will read $(\tilde{\cal H}_{\mbox{\scr lin}} + \tilde{\cal
H}_{\mbox{\scr rest}} )\tilde{\psi}
= E \tilde{\psi}$, with $\tilde{\cal H}_{\mbox{\scr lin}}$ the
leading linearized term 
\begin{equation}
\label{bdglin}
\tilde{\cal H}_{\mbox{\scr lin}} = \left(
\begin{array}{cc}
v_F (p_y -\frac{e}{c} A_y) & \frac{1}{p_F} \{p_x, \Delta(\vec{r})\} \\
\frac{1}{p_F} \{p_x, \Delta^*(\vec{r})\} & -v_F (p_y+\frac{e}{c} A_y)
\end{array}
\right),
\end{equation}
where $v_F = p_F /m$ is the Fermi velocity
and $\tilde{\cal H}_{\mbox{\scr rest}}$ is the remaining piece,
which we will disregard being smaller by ${\cal O}\left(\frac{1}{k_F
d}\right)$.  
Notice that the linearized eigenvalue problem $\tilde{\cal
H}_{\mbox{\scr lin}}
\tilde{\psi} = E \tilde{\psi}$ is gauge covariant, so the linearized
spectrum will be gauge invariant.

\section{Franz--Te\v{s}anovi\'{c} gauge transformation}
\begin{figure}[t]
\noindent
\epsfxsize=8.5cm
\epsffile{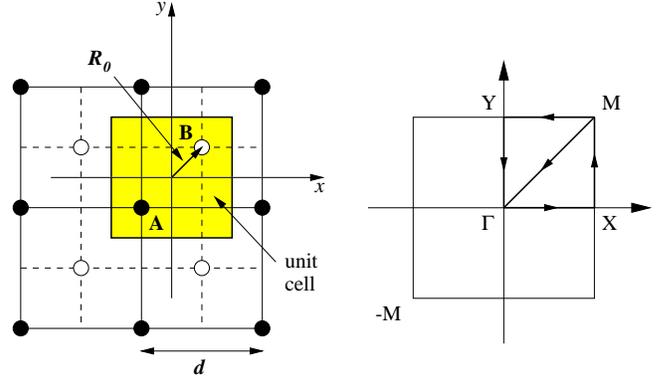}
\caption{(a) A and B sublattices and vortex lattice unit cell. (b) The
corresponding magnetic Brillouin zone.}
\label{figAB}
\end{figure}

Following Franz and Te\v{s}anovi\'{c} \cite{franz00} we can eliminate the
phase factor $e^{i \phi(\vecscr{r})}$ from the off-diagonal components of
the Bogoliubov--de Gennes equation performing the singular gauge
transformation
\begin{eqnarray} \nonumber
{\cal H}_{\mbox{\scr lin}} &\to& \tilde{\cal H}_{\mbox{\scr lin}} =
U^{-1} {\cal H}_{\mbox{\scr lin}} U, \\  U &=& \left(
\begin{array}{cc}
e^{i \phi_A(\vecscr{r})} & 0 \\
0 & e^{-i \phi_B(\vecscr{r})}
\end{array}
\right)
\end{eqnarray}
where $\phi(\vec{r}) = \phi_A(\vec{r}) +\ \phi_B(\vec{r})$. $\phi_A(\vec{r})$ 
and $\phi_B(\vec{r})$ are the contributions to the phase coming from the vortex
sublattices $A$ and $B$ respectively, as defined in Fig.~\ref{figAB} and will 
be computed later in the paper.

The gauge transformed Bogoliubov--de Gennes operator is 
\begin{eqnarray} \nonumber
\tilde{\cal H}_{\mbox{\scr lin}} &=& \left(
\begin{array}{cc}
v_F p_y & v_\Delta p_x \\
v_\Delta p_x & - v_F p_y
\end{array}
\right) \\ &+& m \left(
\begin{array}{cc}
v_F v^A_{sy} & \frac{v_\Delta}{2} (v^A_{sx}-v^B_{sx}) \\
\frac{v_\Delta}{2} (v^A_{sx}-v^B_{sx}) & v_F v^B_{sy}  
\end{array}
\right),
\label{Htransf}
\end{eqnarray}
where $v_\Delta = \Delta_0/p_F$ and the superfluid velocities
corresponding to the $A$ and $B$ sublattices are defined as
\begin{equation}
\vec{v}^\mu_s = \frac{1}{m} (\hbar \nabla \phi_\mu - \frac{e}{c}
\vec{A}), \ \  \mu = A, B.
\end{equation}
The operator (\ref{Htransf}) describes the dynamics of a free Dirac
particle in a periodic potential. We can take advantage of the periodicity
of the potential rewriting our spinors in Bloch form
\begin{equation}
\left(
\begin{array}{c}
u_{\vecscr{k}} \\ v_{\vecscr{k}}
\end{array} \right) = e^{i \vecscr{k}\cdot \vecscr{r}} \left(
\begin{array}{c}
U_{\vecscr{k}} \\ V_{\vecscr{k}}
\end{array} \right)
\end{equation}
where the functions $U_{\vecscr{k}}(\vec{r})$ and
$V_{\vecscr{k}}(\vec{r})$ are themselves periodic on the unit cell
shown in Fig.\ \ref{figAB} and the effective Hamiltonian acting on
the Bloch spinors $(U,V)^T$ is
\begin{eqnarray} \nonumber
{\cal H} &=& \left(
\begin{array}{cc}
v_F (p_y + \hbar k_y) & v_\Delta (p_x + \hbar k_x) \\
v_\Delta (p_x + \hbar k_x) & - v_F (p_y + \hbar k_y)
\end{array}
\right) \\ &+& m \left(
\begin{array}{cc}
v_F v^A_{sy} & \frac{v_\Delta}{2} (v^A_{sx}-v^B_{sx}) \\
\frac{v_\Delta}{2} (v^A_{sx}-v^B_{sx}) & v_F v^B_{sy}  
\end{array}
\right).
\label{HBloch}
\end{eqnarray}

\section{Order parameter spatial distribution}
\label{sect:delta}
The last ingredient we need to determine the quasiparticle spectrum is the 
order parameter spatial distribution in the vortex lattice. The
Bogoliubov--de Gennes method should in principle be used as a set of
equations to be solved self-consistently, thus finding the
quasiparticle spectrum and wavefunctions, and the spatially varying order
parameter distribution. As stated in the introduction, however, we
restrict ourselves
to the case where the vortex core size is small compared to the spacing between
vortices.  Thus, we take the magnitude of the order parameter $\Delta_0$ to be
constant everywhere except at the vortex sites, where it vanishes.  
We also assume that the magnetic field is constant in the sample. The phase 
$\phi(\vec{r})$  of the gap function
$\Delta(\vec{r}) = \Delta_0 e^{i \phi(\vecscr{r})}$ is then obtained by minimizing a
simplified Ginzburg-Landau free energy functional\cite{feder97},    
which has the form
\begin{equation}
F = \mbox{const} \times \int d^2 r\, \left|\nabla \phi(\vec{r}) - 
\frac{2e}{\hbar c} \vec{A}(\vec{r})\right|^2 
\label{free_en}
\end{equation}
where  we have taken $e$ to be negative. 
The phase of the order parameter in the Landau gauge $A_x = -B y$, $ A_y = 0$ 
is then given by
\begin{equation}
\nabla \phi(\vec{r}) =  \lim_{\gamma \to 0} \sum_{\alpha} \frac{\hat{z}\times
(\vec{r}-\vec{r}_\alpha)}{|\vec{r}-\vec{r}_\alpha|^2} e^{-\gamma y^2_{\alpha}}
\end{equation}
where the sum runs over the vortex lattice.
This series can be evaluated in closed form and the final result for the phase
of the order parameter corresponding to the two sublattices $\phi_A(\vec{r})$ 
and $\phi_B(\vec{r})$ for a square vortex lattice with intervortex distance
$2 R_0 = d \sqrt{2}/2$ is
\begin{eqnarray}
\phi_A(\vec{r}) &=& \arg\left[\frac{\theta_1\left(\frac{x+i y}{d}+
\frac{1+i}{\sqrt{2}}\frac{R_0}{d},i\right)}{\theta_1^\prime(0,i)} \right] \\
\phi_B(\vec{r}) &=& \arg\left[\frac{\theta_1\left(\frac{x+i y}{d}-
\frac{1+i}{\sqrt{2}}\frac{R_0}{d},i\right)}{\theta_1^\prime(0,i)} \right]
\end{eqnarray}
where $\theta_1(z,\tau)$ is the antisymmetric elliptic theta function
\cite{chandra85} 
and the modular parameter $\tau = i$ for a square lattice.
These functions are not periodic, as they are gauge dependent, but using the 
properties of the theta functions under translations \cite{chandra85} it is 
possible to show
that the superfluid velocities $\vec{v}^{A,B}_{s}(\vec{r}) = m^{-1} [\hbar
\nabla \phi_{A,B}(\vec{r})-(e/c)A]$ are. For a general lattice, with
two vortices per unit cell, the Fourier representation of
$\vec{v}_s^{\mu}(\vec{r})$ is given by
\begin{equation}
\vec{v}^{\mu}_s(\vec{r}) = \frac{2 \pi \hbar}{m d^2} \sum_{\vecscr{Q} \neq
0} \frac{i \vec{Q} \times \hat{z}}{Q^2} e^{i \vecscr{Q} \cdot
(\vecscr{r}-\vecscr{R}^\mu_0)}, \ \ \mu = A, B
\end{equation} 
where $\vec{R}^B_0 = \vec{R}_0$ and $\vec{R}^A_0 = -\vec{R}_0$,
$\vec{Q}$ are the reciprocal lattice vectors, and $d^2$ is the area of
the magnetic unit cell.
This in turn implies that the total superfluid velocity $\vec{v}_s = m^{-1} [
(\hbar/2) \nabla \phi - (e/c) \vec{A}]$ is also a periodic function
over the vortex lattice as is every gauge invariant quantity derived from it.

\section{Particle-hole symmetry}
\label{sect:symm}
Several symmetries play an important role in this problem, both conceptually
and computationally. Of course, the translational symmetry of the
vortex lattice allowed us to introduce Bloch functions and recast the
calculation of the quasiparticle spectrum into a band theory
framework. 

There are further symmetries that provide some help 
in understanding general features of the spectrum and simplify
the diagonalization of the Bogoliubov--de Gennes equation
(\ref{HBloch}). In the Introduction we mentioned that particle-hole
symmetry and the Bravais nature of the vortex lattice 
are the two key ingredients that lead to the gaplessness of the
quasiparticle spectrum. The importance of the Bravais
lattice will be emphasized in detail in
Sections~\ref{sect:pertgamma}-\ref{sect:epsilon} where the band
structure is studied in perturbation theory. Here we want to
focus on particle-hole symmetry. If $(U_{n
\vecscr{k}}(\vec{r}),V_{n\vecscr{k}}(\vec{r}))^T$ is an eigenvector of
the Bogoliubov--de Gennes equation (\ref{HBloch}) with eigenvalue $E_{n
\vecscr{k}}$ where $n$ is a band index and $\vec{k}$ is a wave vector
in the first Brillouin zone, define
\begin{equation}
\begin{array}{c}
\tilde{U}_{n \vecscr{k}}(\vec{r}) = - V^*_{n \vecscr{k}}(-\vec{r}) \\
\tilde{V}_{n \vecscr{k}}(\vec{r}) = U^*_{n \vecscr{k}}(-\vec{r}).
\end{array}
\label{spintransf}
\end{equation}
We claim that the spinor $(\tilde{U}_{n \vecscr{k}},\tilde{V}_{n
\vecscr{k}})^T$ is an eigenvector of the Bogoliubov--de Gennes operator
(\ref{HBloch}) with eigenvalue $-E_{n \vecscr{k}}$. It is important to
notice that in this way we are proving that particle-hole symmetry
doesn't just hold on the whole spectrum as a set, it is an exact
symmetry of the linearized Bogoliubov--de Gennes operator at
\emph{every} point in the Brillouin zone. In particular this means
that the entire band structure should be exactly particle-hole
symmetric. The proof of this statement is most easily constructed
rewriting the transformation (\ref{spintransf}) more explicitly in an
operator notation 
\begin{equation}
\left( \begin{array}{c}
\tilde{U}_{n \vecscr{k}} \\
\tilde{V}_{n \vecscr{k}}
\end{array} \right) = {\cal S}\left( 
\begin{array}{c}
U_{n \vecscr{k}} \\
V_{n \vecscr{k}}
\end{array}    \right)
\end{equation}
with $\cal S = C R T$ where $\cal C$ is the complex conjugation
operator, $\cal R$ is the reflection through the origin operator and
${\cal T} = - i \sigma_2$. It is easy to show that the linearized
Bogoliubov--de Gennes operator ${\cal H}$, defined in equation
(\ref{HBloch}), and $\cal S$ anticommute $\{{\cal S},{\cal H}\} = 0$
using the property that the superfluid velocities for the two
sublattices $A$ and $B$ are related by
\begin{equation}
\vec{v}^B(\vec{r})=-\vec{v}^A(-\vec{r}),
\label{vtransf}
\end{equation} 
in the coordinate system
sketched in Fig.~\ref{figAB}, and vice versa. 
The particle-hole symmetry of the band structure is also observed
explicitly in the 
numerical spectra as can be seen in Figs.~\ref{figdelta1}-\ref{figdelta4}
for the case of a Bravais lattice. In the proof above, we only used
the existence of a center of inversion in the vortex lattice. This
holds also in the case of a non-Bravais lattice with two vortices per
unit cell and so we expect to find a particle-hole symmetric band
structure as well, which agrees with the numerical results shown in
Fig.~\ref{figp}. However, the proof fails, in general, for more complicated
structures, with more than two vortices per unit cell where there
is not a center of inversion symmetry anymore.

\section{Band structure calculations through numerical diagonalization}
\label{sect:numband}
We have run extensive numerical diagonalization of the Hamiltonian
(\ref{Hdirac}), scanning the magnetic Brillouin zone for different
values of the anisotropy ratio $\alpha_D= v_F/v_\Delta$.

Unlike Franz and Te\v{s}anovi\'{c}, we decided to run our band
structure calculations in real space, rather than momentum space.
The real space discretization leads to a sparse
representation of the Hamiltonian which allows us to look at finer
meshes then we would be able to in reciprocal space.  The algorithm we
used for finding the eigenfunctions and eigenvalues 
is a modified Lanczos-type method called the implicitly restarted
Arnoldi method \cite{sorensen95}, implemented through the public
domain Fortran 77 package ARPACK.

We note that the superfluid velocity $\vec{v}_s$ is singular at the vortex
points (because we are
taking the limit where the coherence length goes to zero), and its
Fourier components (\ref{pert}) decay only algebraically. The real space and
reciprocal space methods differ in the way in which they treat the 
large-wavevector cutoff. We think it
is more direct to control the effects of this singularity in real
space than in reciprocal space. However, we noticed a strong
sensitivity of the results to the position of the vortex with respect
to the real space mesh, until we cut the singularity off with a
gaussian smoothing factor on a scale of a few grid points.
Using a real space approach, the discretization of the superfluid
velocity and it's regularization near the vortices are controlled
by separate parameters and can be optimized independently.

A disadvantage of using a real space approach for fermions is that
one has to deal with the fermion doubling problem \cite{kogut83}. 
If one discretizes the Dirac Hamiltonian in two spatial dimensions in
the most straightforward way using a rectangular mesh then one finds
(in the absence of a scalar or vector potential) that there are 3
spurious low-energy modes at the boundaries of the Brillouin zone of
the mesh, i.e. when $\vec{k}/\pi = (a_x^{-1},0), (0,a_y^{-1})$ or
$(a_x^{-1},a_y^{-1})$, where $a_x$ and $a_y$ are the mesh
spacings. This is a problem because the spurious modes get mixed with
the physical low-energy modes (near $\vec{k}=0$) by the inhomogeneous
potential. This problem has been thoroughly investigated in the lattice
gauge theory community and one way of getting rid of it was introduced
by Wilson \cite{wilson75}. Essentially, one introduces a $k-$dependent
mass term in the Dirac Hamiltonian that vanishes like $k^2$ at the center of
the Brillouin zone and lifts the zero-energy modes at the boundaries
of the unit cell to some high energy scale. Our choice for the Wilson
term is
\begin{eqnarray} \nonumber
{\cal H}_W &=& \hbar v_F \left(\lambda_x \frac{1-\cos k_x
a_x}{a_x}+\lambda_y \frac{1-\cos k_y a_y}{a_y}\right) \sigma_2 \\ &=&
\hbar v_F \left(\lambda_x a_x \delta_x^2
+ \lambda_y a_y \delta_y^2 \right) \sigma_2,
\label{Hwilson}
\end{eqnarray}
where $\delta_x^2$ and $\delta_y^2$ are the second difference
operators on the lattice. We are interested here in wavefunctions
which vary smoothly on the scale of the vortex lattice spacing $d$. If
we choose $a_x$ and $a_y$ sufficiently small compared to $d$, for
fixed $\lambda_x$ and $\lambda_y$ the Wilson term will have negligible
effect on the physical states, but the spurious states, which
oscillate on the scale of $a$, will be pushed up to very high energies.

\begin{figure}[t]
\noindent
\epsfxsize=8.5cm
\epsffile{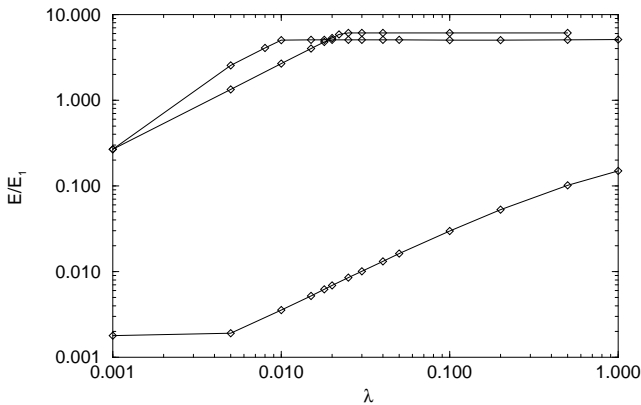}
\caption{Scaling analysis for the lowest three distinct eigenvalues as
a function of the Wilson parameter $\lambda$ in the isotropic case
$\alpha_D = v_F/v_\Delta = 1$ ($\lambda_x = \lambda_y$) for fixed
number of mesh points $N_x = N_y = 70$ per unit cell. Energies are in units of
$E_1 = \hbar v_F /d$.}
\label{figscaling}
\end{figure}

When the superfluid velocities $\vec{v}_s^{\mu}(\vec{r})$ are included,
the Wilson scheme breaks the symmetry that keeps the spectrum gapless
at the center of the Brillouin zone of the vortex lattice. Therefore,
we have to perform a finite size scaling analysis to 
determine whether the small gaps we see in the numerics
disappear in the limit $\lambda_x,\lambda_y \to 0$. In
Fig.~\ref{figscaling} one can see an example of such an analysis for the
isotropic case $\alpha_D =1$. If the spectrum is gapless one expects the
Wilson term to open a gap linear in $\lambda_x$ and $\lambda_y$, which
is very close to what we see for the lowest eigenvalue. We use this
analysis to choose a value of $\lambda_x$ and
$\lambda_y$ that is as small as possible but that doesn't get us in
the region where we can see effects of the fermion doubling problem.

\begin{figure}[t]
\noindent
\epsfxsize=8.5cm
\epsffile{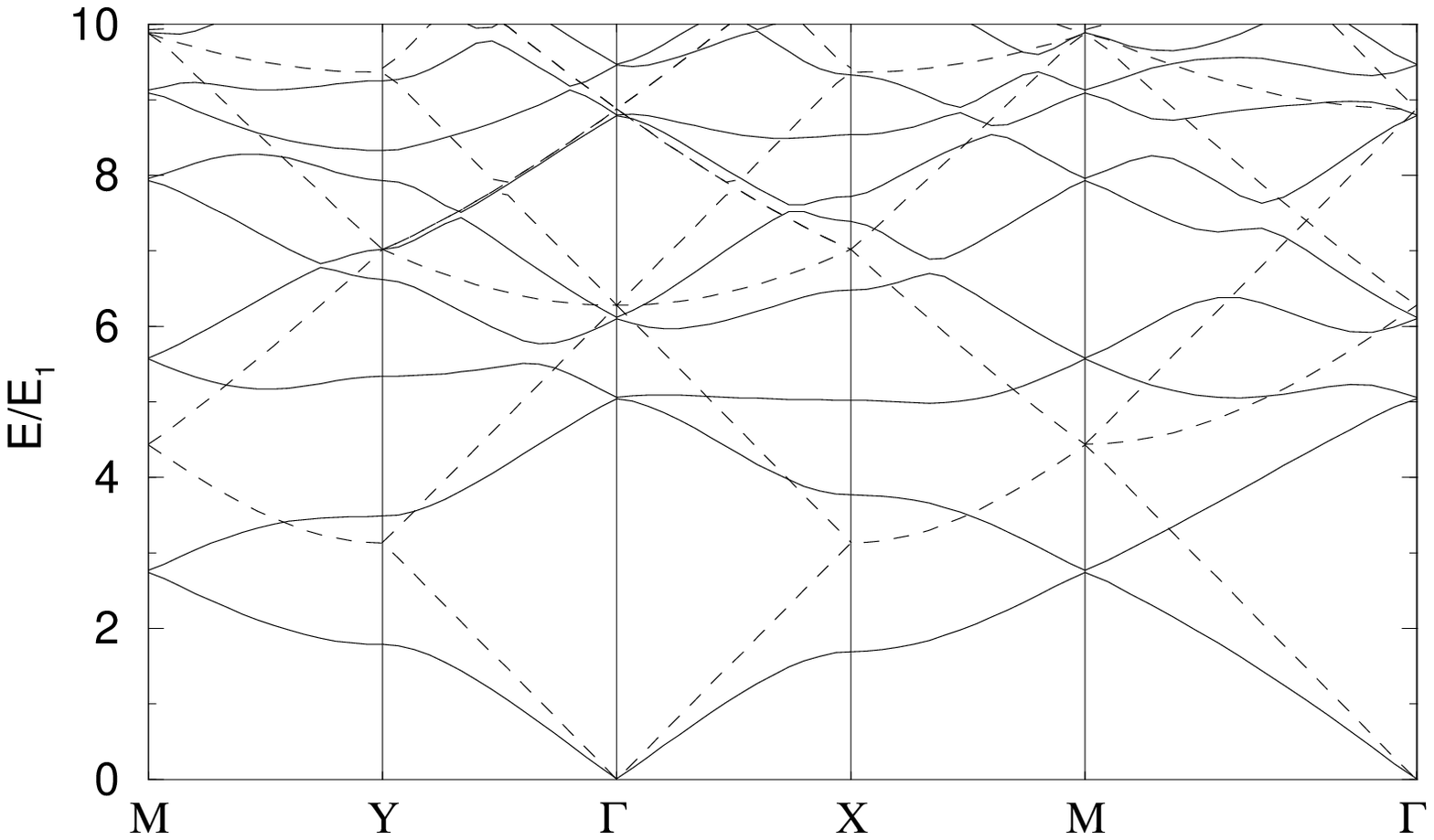}
\\*
\noindent
\epsfxsize=8.5cm
\epsffile{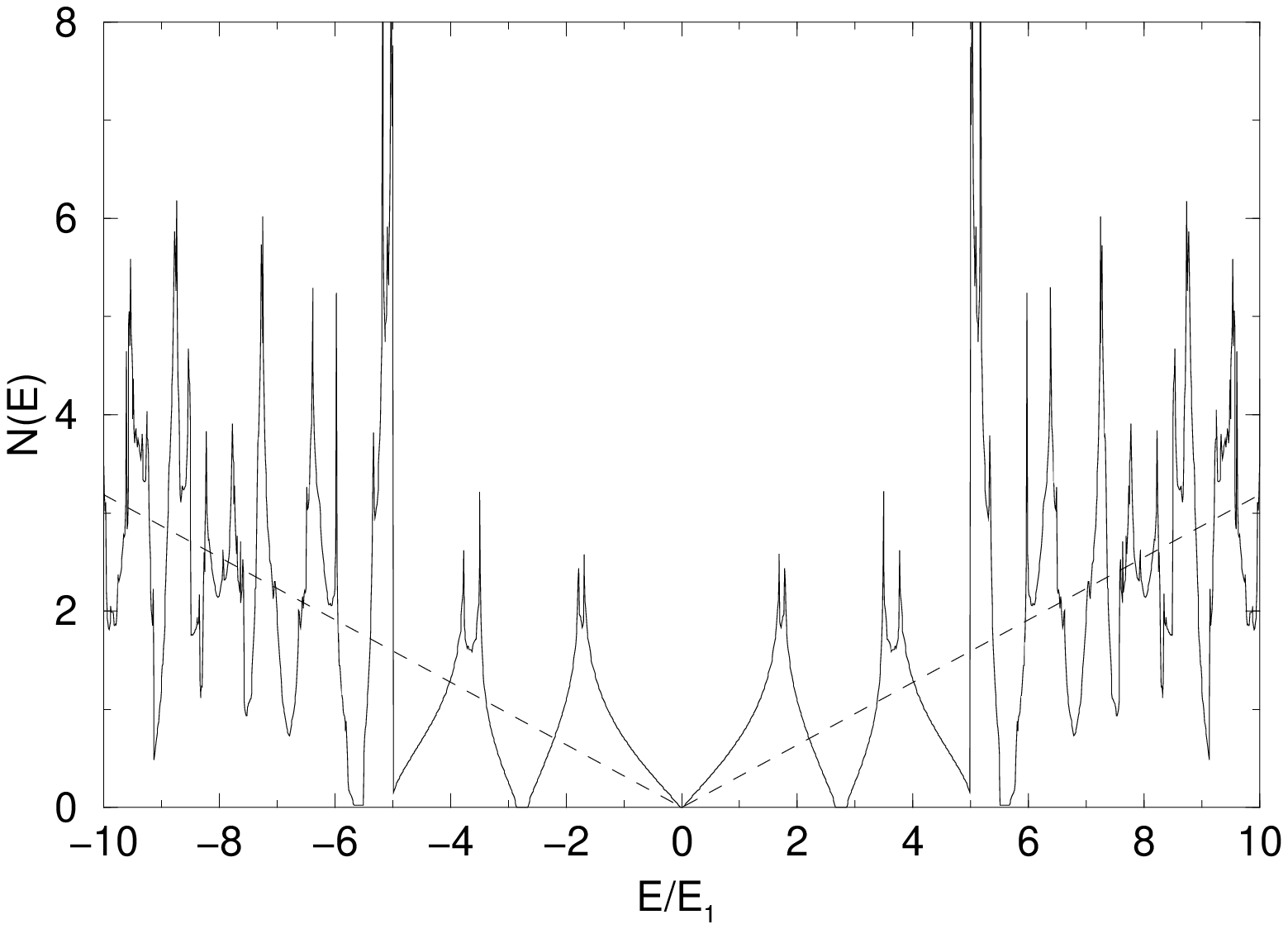}
\caption{Band structure and density of states for a square lattice
with $\alpha_D = v_F /v_\Delta = 1$. The dashed line is the spectrum
of the free Dirac Hamiltonian. Only the positive energy bands 
are plotted for clarity, the negative energy ones can 
be obtained by particle-hole symmetry (the density of states plot shows the 
overall particle-hole symmetry explicitly).
Energies are in units of $E_1=\hbar v_F/d$.}
\label{figdelta1}
\end{figure}

\begin{figure}[t]
\noindent
\epsfxsize=8.5cm
\epsffile{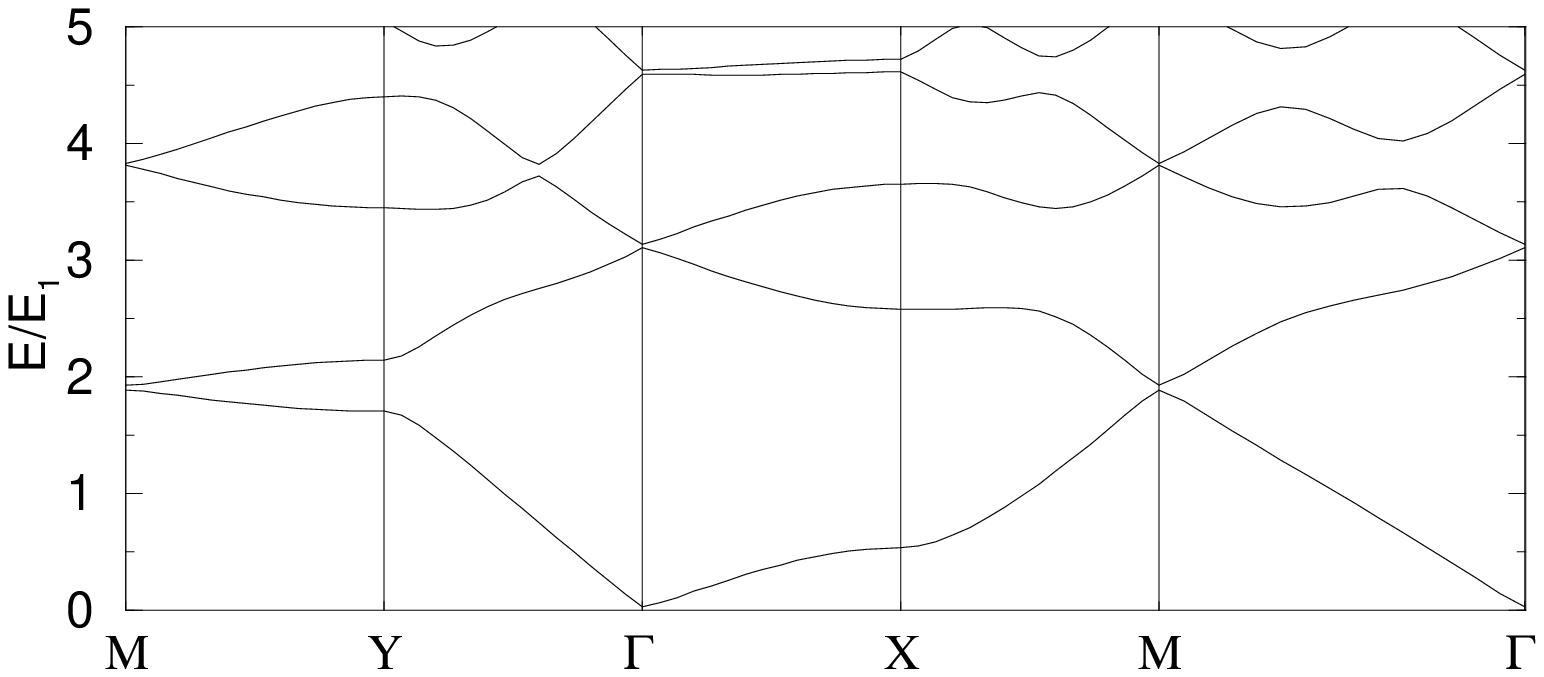}
\\*
\noindent
\epsfxsize=8.5cm
\epsffile{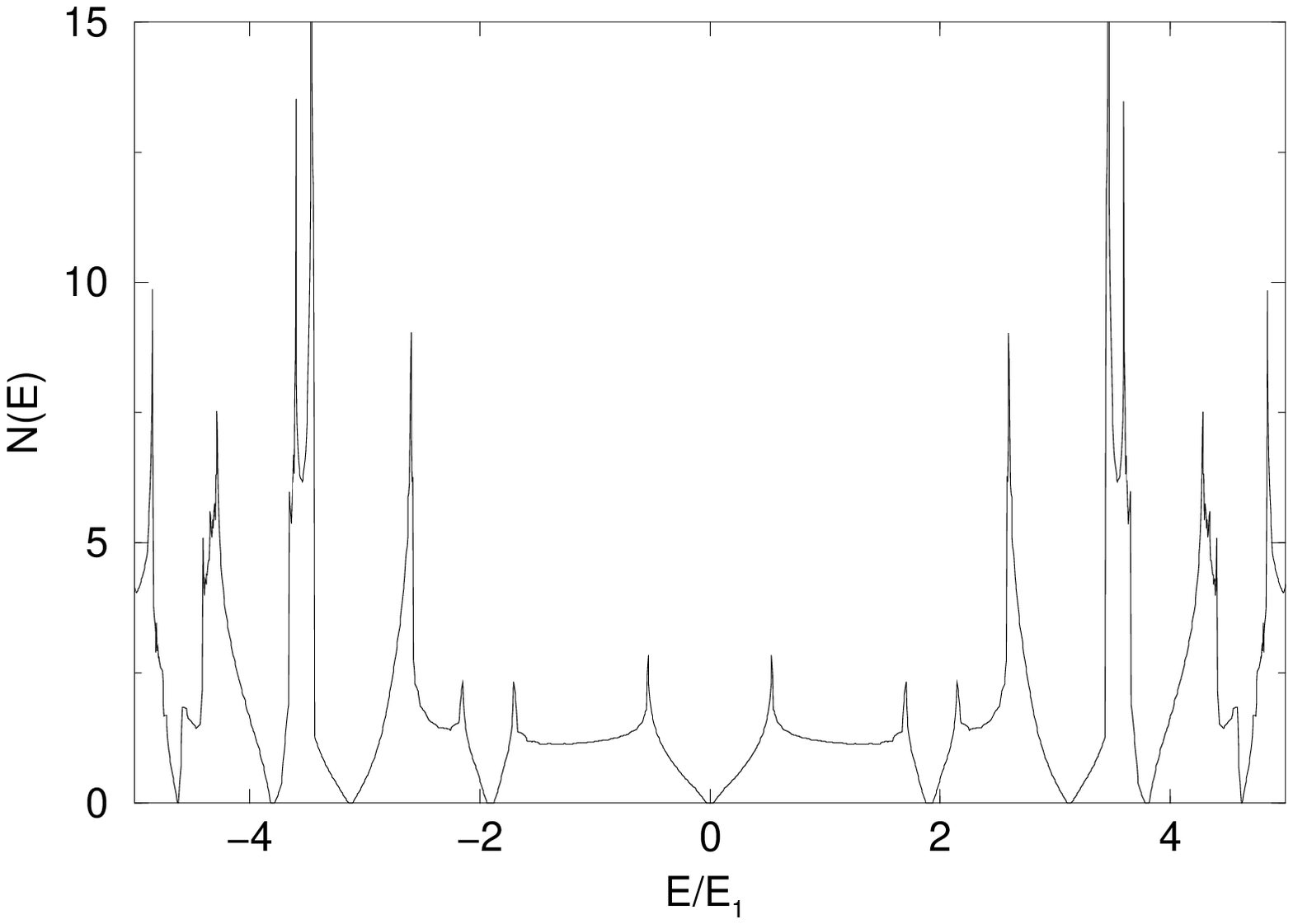}
\caption{Band structure and density of states for a square lattice
with $\alpha_D = v_F/ v_\Delta = 2$. Negative energy bands can 
be obtained by particle-hole symmetry.}
\label{figdelta2}
\end{figure}

\begin{figure}[t]
\noindent
\epsfxsize=8.5cm
\epsffile{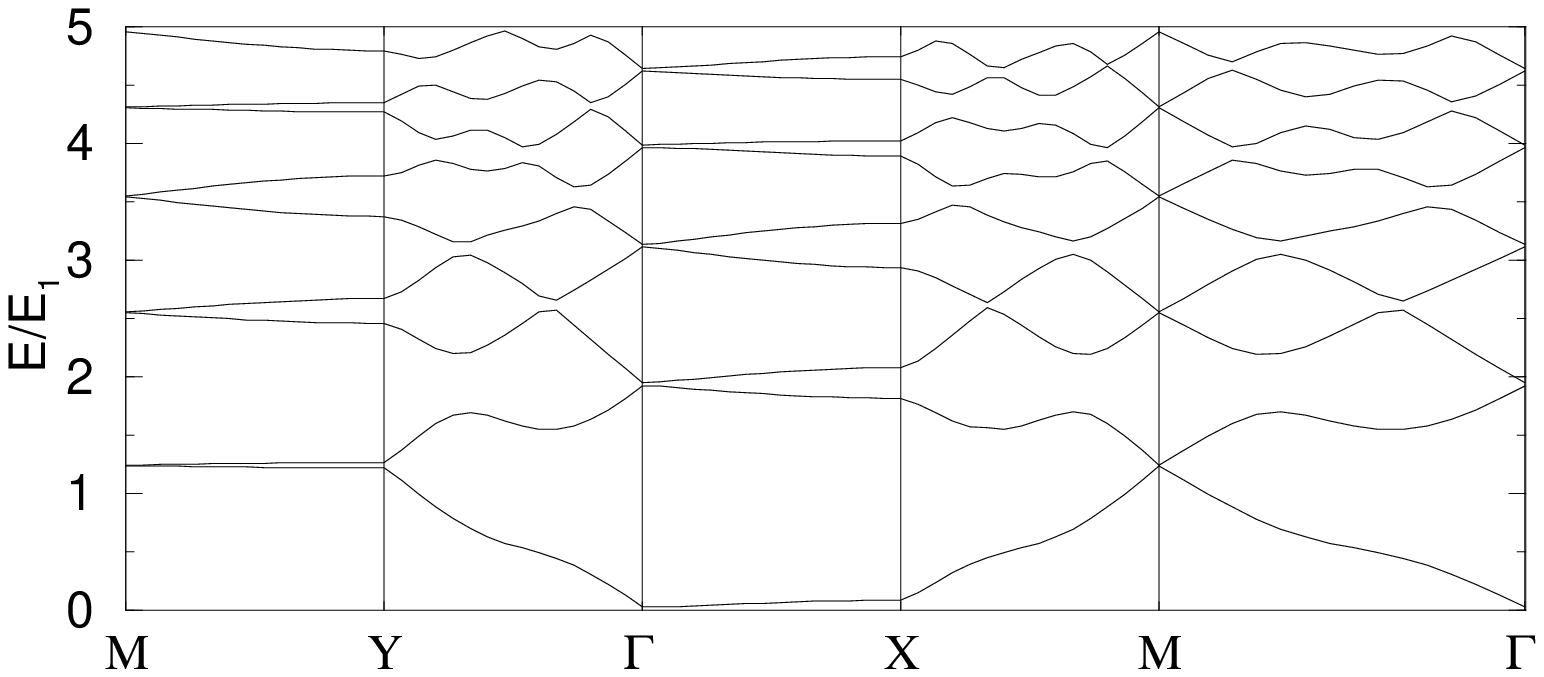}
\\*
\noindent
\epsfxsize=8.5cm
\epsffile{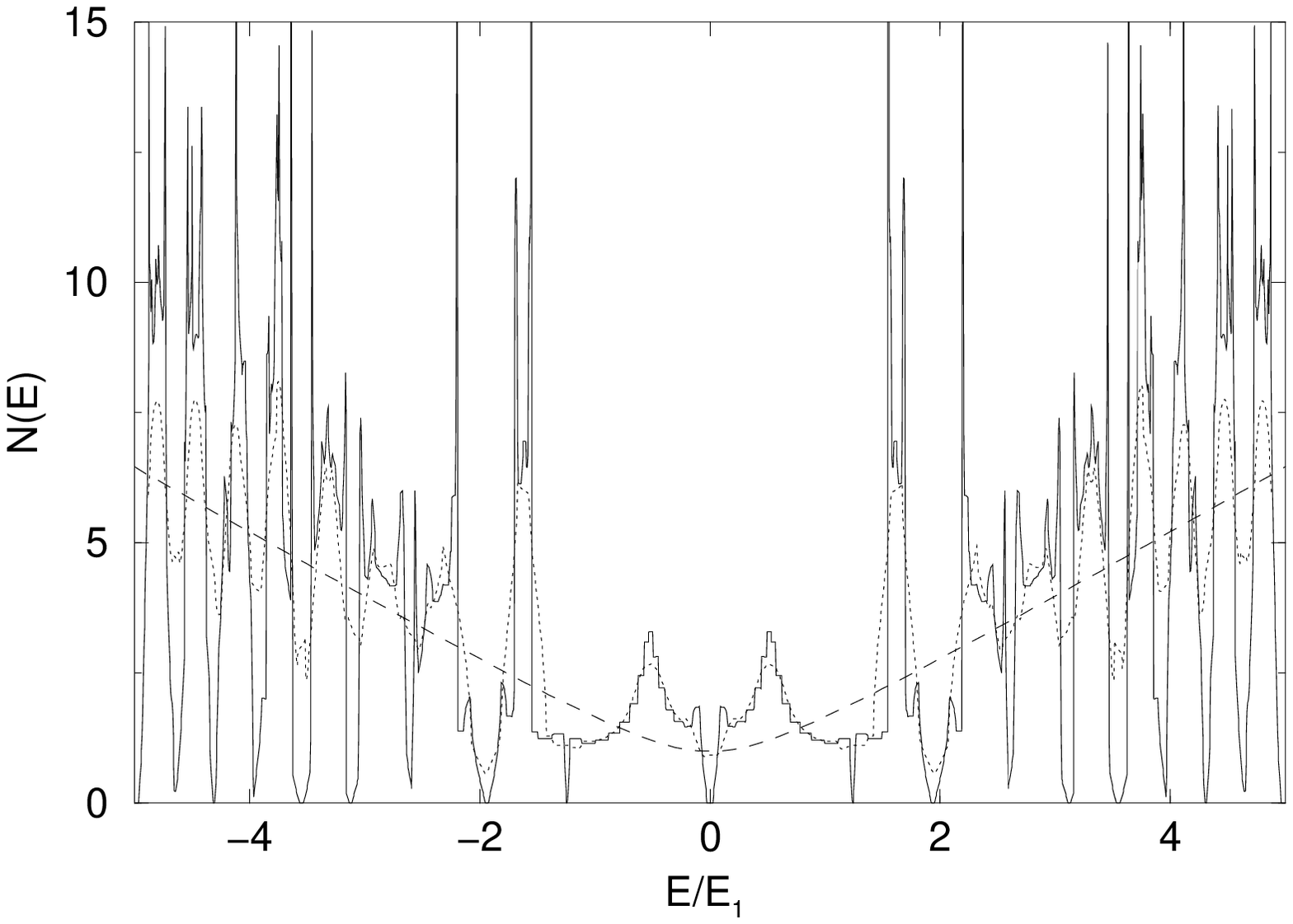}
\caption{Band structure and density of states for a square lattice
with $\alpha_D = v_F / v_\Delta = 4$. In figure (b), the dotted line
is a smoothed out density of states on a scale $\Delta\!E = 0.25 \hbar v_F/d$,
while the dashed line is the semiclassical density of
states for a square lattice and $\alpha_D = 4$.  
Negative energy bands can 
be obtained by particle-hole symmetry.} 
\label{figdelta4}
\end{figure}

We have computed the band structure for several values of the
anisotropy ratio $\alpha_D$. Figs.~\ref{figdelta1}-\ref{figdelta4}
show the band structure along symmetry lines and the density of states
for the case of a square lattice with anisotropy $\alpha_D=1$, $2$ and
$4$ respectively. The quasiparticle energy bands in a square vortex
lattice are symmetric under the exchange of $k_x \to -k_x$
or $k_y \to -k_y$ and so only positive $k_x$ and $k_y$ have been
considered. As we have already discussed in Section \ref{sect:symm},
the bands are particle-hole symmetric, therefore only positive energy
bands are plotted. 

Noticeable features are the absence of a gap at the
$\Gamma$ point and further band crossings at higher energies also at
the $\Gamma$ point. As we mentioned earlier and as will be analyzed in
more detail in the following sections, the Bravais nature of the
vortex lattice plays an essential role in keeping the spectrum
gapless. The M point is also a special point, there are band crossings
although not at zero energy. Also these band crossings will become
avoided crossings once we modify the lattice into a non-Bravais one. 

\section{Perturbation theory at the $\Gamma$ point}
\label{sect:pertgamma}
One of the advantages of the Franz--Te\v{s}anovi\'{c} gauge
transformation is that it rephrases the problem in a form that is well
suited to a perturbative analysis. The original linearized
Bogoliubov--de Gennes equation doesn't easily separate into an
exactly solvable unperturbed part plus a periodic (or quasi-periodic)
perturbation while (\ref{HBloch}) is immediately recognizable as a
two-dimensional free Dirac Hamiltonian perturbed by an effective
periodic vector and scalar potential
\begin{eqnarray}
{\cal H} &=& {\cal H}_0 + {\cal H}_1 \label{Hdirac} \\
 {\cal H}_0 &=&  v_F (p_y+\hbar k_y) \sigma_3 + v_\Delta (p_x + \hbar k_x)
\sigma_1 \\
{\cal H}_1 &=& \frac{m}{2}\left[ v_F \left(v_{sy}^A+v_{sy}^B\right)
\sigma_0 + v_F \left(v_{sy}^A-v_{sy}^B\right) \sigma_3
\right. \nonumber 
\\ & & \left. ~ ~ ~ ~ ~ ~ ~ +  v_\Delta
\left( v_{sx}^A-v_{sx}^B\right) \sigma_1 \right]
\end{eqnarray}   
where $\sigma_i$, $i=1,\ldots,3$ are the Pauli spin matrices and
$\sigma_0$ is the 2 by 2 identity matrix.

The real space unit cell has to enclose an even number of vortices as
each one of them carries only half a flux quantum $\Phi_0 = h c/|e|$.
We will consider only unit cells with two vortices, but we will not
restrict our analysis to Bravais lattices. The origin will be chosen
at the center of inversion symmetry  as shown in
Fig.\ \ref{figAB}, and the position of the two vortices $\vec{R}_0$
and $-\vec{R}_0$  is left arbitrary. The question we will try to address in
perturbation theory is whether the presence of vortices introduces a
gap in the quasiparticle spectrum. Then, if ${\cal H}_1$ is sufficiently
small, we can limit our
calculation to the center of the magnetic Brillouin zone (the $\Gamma$
point). In Fourier space, the Hamiltonian (\ref{Hdirac}) at the
$\Gamma$ point can be written as
\begin{eqnarray}
{\cal H}_{\vecscr{p}_1 \vecscr{p}_2} = \left(v_F p_{1x} \sigma_1 +
v_\Delta p_{1y} \sigma_3\right) \delta_{\vecscr{p}_1 \vecscr{p}_2} +
{\cal V}^{(0)}_{\vecscr{p}_1-\vecscr{p}_2} \sigma_0  \nonumber \\ +
{\cal V}^{(3)}_{\vecscr{p}_1-\vecscr{p}_2} \sigma_3 +
{\cal V}^{(1)}_{\vecscr{p}_1-\vecscr{p}_2} \sigma_1 
\end{eqnarray}
where the periodic potentials are
\begin{eqnarray}
{\cal V}^{(0)}_{\vecscr{Q}} &=& 2 \pi \frac{\hbar v_F}{d^2} \frac{i Q_x}{Q^2}
\cos \vec{Q}\cdot \vec{R}_0 \nonumber \\
{\cal V}^{(1)}_{\vecscr{Q}} &=& - 2 \pi \frac{\hbar v_\Delta}{d^2} \frac{Q_y}{Q^2}
\sin \vec{Q}\cdot\vec{R}_0 \label{pert} \\
{\cal V}^{(3)}_{\vecscr{Q}} &=& 2 \pi \frac{\hbar v_F}{d^2} \frac{Q_x}{Q^2}
\sin \vec{Q} \cdot \vec{R}_0. \nonumber
\end{eqnarray}
The unperturbed eigenvalues are $E^{(0)}_{\vecscr{Q},\pm} = \pm
\sqrt{v_F^2 \hbar^2 Q_y^2+ v_\Delta^2 \hbar^2 Q_x^2}$ where $\vec{Q}
= m \vec{b}_1 + n \vec{b}_2$ are reciprocal lattice
vectors. The basis vectors $\vec{b}_1$ and $\vec{b}_2$ are chosen
such that, if $\vec{a}_1$ and $\vec{a}_2$ are the basis vectors of the
direct lattice, $\vec{a}_i \cdot \vec{b}_j = 2 \pi \delta_{ij}$. The 
unperturbed eigenvectors can also be computed explicitly. If we write
them in the form 
\begin{equation} 
|E_{\vecscr{Q},\pm}^{(0)}\rangle = \frac{e^{i
\vecscr{Q}\cdot \vecscr{r}}}{d} \left(
\begin{array}{c} 
\alpha_{\vecscr{Q},\pm} \\
\beta_{\vecscr{Q},\pm}
\end{array} \right)
\end{equation}
the coefficients $\alpha$ and $\beta$ can be chosen real and have the
symmetry property 
\begin{equation}
\alpha_{-\vecscr{Q},+} = - \alpha_{\vecscr{Q},-}
\label{alphasym}
\end{equation}
and analogously for $\beta$. 

We are interested in finding if the levels $E^{(0)}_+$ and $E^{(0)}_-$
are split by the perturbing potentials (\ref{pert}). Either even or
odd orders in
perturbation theory have to vanish because if we change the sign
of the superfluid velocity the splitting should remain unchanged as
a consequence of particle-hole symmetry. It's 
easy to see that first order perturbation theory vanishes as the
unperturbed eigenvectors $|E^{(0)}_\pm\rangle$ are constant spinors and
the superfluid velocities $\vec{v}^{A,B}$ have zero average.

To find the effect of the perturbation to second order, we have to
write an  effective Hamiltonian for the two lowest energy bands. Using
the formalism of Brillouin--Wigner perturbation theory, we may write
the Schr\"{o}dinger equation for a given wavevector in the form
\begin{equation}
{\cal H}_{\mbox{\scr eff}}(E) \Psi = E \Psi
\end{equation}
where $\Psi$ is a two-component spinor, and ${\cal H}_{\mbox{\scr
eff}}$ is a 2 by 2 matrix defined by
\begin{equation}
{\cal H}_{\mbox{\scr eff}}(E) = {\cal P}_{\vecscr{k}} \left( {\cal H}_0 + {\cal
H}_1 \frac{1- {\cal P}_{\vecscr{k}}}{E- {\cal H}_0 - {\cal H}_1} {\cal H}_1
\right) {\cal P}_{\vecscr{k}}
\label{Heff}
\end{equation}
where ${\cal P}_{\vecscr{k}}$ is the projection operator onto the 
two-dimensional subspace spanned by the unperturbed eigenvectors
$|E^{(0)}_{\vecscr{k},\pm}\rangle$. Since we are looking at the behavior near
zero energy, we can set $E=0$ in (\ref{Heff}). Also, to second order
in ${\cal H}_1$, we can neglect the ${\cal H}_1$ in the denominator of
the second term. Then, at the $\Gamma$ point, where ${\cal P}_0 {\cal
H}_0 {\cal P}_0 =0$ (we are using the notation ${\cal
P}_{\vecscr{k}=(0,0)} = {\cal P}_0$ at the $\Gamma$ point), we have 
\begin{equation}
{\cal H}_{\mbox{\scr eff}}^{(2)} = - {\cal P}_0 {\cal H}_1 \frac{1-
{\cal P}_0}{{\cal H}_0} {\cal H}_1 {\cal P}_0.
\end{equation} 
The matrix elements of this 2 by 2 matrix can be explicitly calculated
\end{multicols}
\begin{eqnarray}
\langle E^{(0)}_{0,+}|{\cal H}_{\mbox{\scr eff}}^{(2)}|
E^{(0)}_{0,+}\rangle &=& -
\sum_{\vecscr{Q}\neq 0, i = \pm}
\frac{1}{E^{(0)}_{\vecscr{Q},i}} \left| \alpha_{\vecscr{Q},i}
{\cal V}^{(0)}_{\vecscr{Q}} + \alpha_{\vecscr{Q},i}
{\cal V}^{(3)}_{\vecscr{Q}}+\beta_{\vecscr{Q},i}
{\cal V}^{(1)}_{\vecscr{Q}}\right|^2 
\label{pert++} \\
\langle E^{(0)}_{0,+}|{\cal H}_{\mbox{\scr eff}}^{(2)}|
E^{(0)}_{0,-}\rangle &=& 
\sum_{\vecscr{Q}\neq 0, i = \pm}
\frac{1}{E^{(0)}_{\vecscr{Q},i}} \left( \alpha_{\vecscr{Q},i}
{\cal V}^{(0)}_{\vecscr{Q}} + \alpha_{\vecscr{Q},i}
{\cal V}^{(3)}_{\vecscr{Q}}+\beta_{\vecscr{Q},i}
{\cal V}^{(1)}_{\vecscr{Q}}\right) \left(\beta_{\vecscr{Q},i}
{\cal V}^{(0)}_{\vecscr{Q}} + \beta_{\vecscr{Q},i}
{\cal V}^{(3)}_{\vecscr{Q}}-\alpha_{\vecscr{Q},i} {\cal V}^{(1)}_{\vecscr{Q}}
\right) \label{pert+-} \\
\langle E^{(0)}_{0,-}|{\cal H}_{\mbox{\scr
eff}}^{(2)}|E^{(0)}_{0,-}\rangle &=& -
\sum_{\vecscr{Q}\neq 0, i = \pm}
\frac{1}{E^{(0)}_{\vecscr{Q},i}} \left| \beta_{\vecscr{Q},i}
{\cal V}^{(0)}_{\vecscr{Q}} - \beta_{\vecscr{Q},i}
{\cal V}^{(3)}_{\vecscr{Q}}+\alpha_{\vecscr{Q},i}
{\cal V}^{(1)}_{\vecscr{Q}}\right|^2. \label{pert--} 
\end{eqnarray}
\begin{multicols}{2}
Notice that for a Bravais lattice, $\vec{R}_0 =
(\vec{a}_1/4)+(\vec{a}_2/4)$ and so if we write the reciprocal lattice
vectors as $\vec{Q} = m \vec{b}_1+ n \vec{b}_2$, for $m+n$ even
${\cal V}^{(1)}_{\vecscr{Q}} = {\cal V}^{(3)}_{\vecscr{Q}} = 0$ while
for $m+n$ odd  
${\cal V}^{(0)}_{\vecscr{Q}} = 0$. Summing up the
series in (\ref{pert++}-\ref{pert--}) using the above mentioned
property of 
the Fourier components of the potential and the symmetry of the
unperturbed eigenfunctions (\ref{alphasym}) it is easy to show that
${\cal H}_{\mbox{\scr eff}}^{(2)}$ vanishes, i.e. the perturbation
does not open a gap in the spectrum up to second order. 

The third order contribution to the effective Hamiltonian at the
$\Gamma$ point is
\begin{equation}
{\cal H}_{\mbox{\scr eff}}^{(3)} = {\cal P}_0 {\cal H}_1 \frac{1-
{\cal P}_0}{{\cal H}_0} {\cal H}_1 \frac{1-
{\cal P}_0}{{\cal H}_0} {\cal H}_1 {\cal P}_0.
\end{equation}  
We will show that this matrix is also vanishing, to illustrate the
point let's consider the  off-diagonal matrix element
\end{multicols}
\begin{eqnarray}
\langle E^{(0)}_{0,+}|{\cal H}_{\mbox{\scr eff}}^{(3)}|
&& E^{(0)}_{0,-}\rangle = 
\sum_{\vecscr{Q}_1,\vecscr{Q}_2\neq 0, i_1,i_2 = \pm}
\frac{1}{E^{(0)}_{\vecscr{Q}_1,i_1} E^{(0)}_{\vecscr{Q}_2,i_2}} 
\left(\alpha_{\vecscr{Q}_1,i_1}
{\cal V}^{(0)}_{\vecscr{Q}_1} + \alpha_{\vecscr{Q}_1,i_1}
{\cal V}^{(3)}_{\vecscr{Q}_1}+\beta_{\vecscr{Q}_1,i_1}
{\cal V}^{(1)}_{\vecscr{Q}_1}\right) \nonumber \\
&&  \times \left(-\beta_{\vecscr{Q}_2,i_2}
{\cal V}^{(0)}_{\vecscr{Q}_2} - \beta_{\vecscr{Q}_2,i_2}
{\cal V}^{(3)}_{\vecscr{Q}_2}+ \alpha_{\vecscr{Q}_2,i_2} {\cal
V}^{(1)}_{\vecscr{Q}_2 } \right)
\left[{\cal V}^{(0)}_{\vecscr{Q}_2 -
\vecscr{Q}_1}(\alpha_{\vecscr{Q}_1,i_1} \alpha_{\vecscr{Q}_2,i_2} +
\beta_{\vecscr{Q}_1,i_1} \beta_{\vecscr{Q}_2,i_2}) \right. \nonumber \\
&& + \left. {\cal
V}^{(3)}_{\vecscr{Q}_2 - \vecscr{Q}_1} (\alpha_{\vecscr{Q}_1,i_1}
\alpha_{\vecscr{Q}_2,i_2} - \beta_{\vecscr{Q}_1,i_1}
\beta_{\vecscr{Q}_2,i_2}) + {\cal V}^{(1)}_{\vecscr{Q}_2 -
\vecscr{Q}_1} (\alpha_{\vecscr{Q}_1,i_1} \beta_{\vecscr{Q}_2,i_2} +
\beta_{\vecscr{Q}_1,i_1} \alpha_{\vecscr{Q}_2,i_2}) \right] 
\label{third}
\end{eqnarray}
\begin{multicols}{2} 

Once again, we will use the Bravais lattice symmetry and consider
pairs of terms in the sum corresponding to $(\vec{Q},\pm)$ and
$(-\vec{Q}, \mp)$. In particular, if we write $\vec{Q}_{1,2} = m_{1,2}
\vec{b}_1 + n_{1,2} \vec{b}_2$ and look at the relative sign
of the terms in the sum (\ref{third}) changing the signs of
$\vec{Q}_1$ and $\vec{Q}_2$ simultaneously, we find:

\begin{center}
\begin{tabular}{ccc}
$\vec{Q}_1$ & $\vec{Q}_2$ & relative sign \\ [1.0ex]
even & even & $-$ \\
even & odd & $-$ \\
odd & even & $-$ \\
odd & odd & $-$
\end{tabular}
\end{center}

where ``even'' and ``odd'' refer to the parity of
$m_{1,2}+n_{1,2}$. Adding up all the terms in pairs, we see that this
matrix element vanishes as well as the diagonal ones, as can be
easily checked.  This calculation can be immediately extended to the
fourth order, using exactly the same arguments and in fact, by
induction, to any
other order to show that to every order in perturbation theory the
effective Hamiltonian at the center of the Brillouin zone for the lowest two
bands vanishes. We thus find that to every order in perturbation
theory the potential ${\cal H}_1$ does not open a gap in the spectrum.

\section{Perturbation theory away from the $\Gamma$ point}
\label{sect:pertgen}
The above analysis can be extended away from the $\Gamma$ point. In
particular we are interested in determining whether it's possible to
find other points (possibly not symmetry points) in the
Brillouin zone where the spectrum is gapless. In the following, we
will specialize our analysis to the case of a square lattice. Based
on their numerical analysis, Franz and
Te\v{s}anovi\'{c} \cite{franz00} claim that for large enough anisotropy
there is a whole \emph{line} of zeroes that develops, in our notation, along a
line parallel to the
$k_x$ axis at a value of $k_y$ which depends on the anisotropy. 
(Note that our convention for the $x$ and $y$ axis is the
opposite of Franz and Te\v{s}anovi\'{c}.) However, purely on symmetry grounds,
our effective Hamiltonian for the lowest two bands should be a complex
hermitian 2 by 2 matrix. 
Particle-hole symmetry restricts the number of independent components to
three (the effective Hamiltonian has to be traceless at every
point in $k-$space) but being in two dimensions we only have two
parameters $k_x$ and $k_y$ to vary. The system is obviously
overdetermined and for a generic Hamiltonian of this kind we would not
expect any zeroes, let alone lines of zeroes. The only way in which
zeroes in the spectrum can develop is through some extra symmetry of
the problem. We will see that there is such a symmetry only along the
$k_y =0$ axis.

For a general wavevector $\vec{k}$, at energy $E=0$, we can write the
effective Hamiltonian (\ref{Heff}) in the form 
\begin{equation}
{\cal H}_{\mbox{\scr eff}} = A(\vec{k}) \sigma_3 + B(\vec{k}) \sigma_1
+ C(\vec{k}) \sigma_2,
\end{equation}
where $\sigma_1,\sigma_2$ and $\sigma_3$ are the Pauli spin matrices
and $A, B$ and $C$ are real functions.

In order for zero-energy states to exist $A, B$ and $C$ must all
vanish simultaneously. We have seen that this happens at the $\Gamma$
point, for a Bravais vortex lattice. To see if that can happen at
other points, we will first consider the symmetry line $k_x = 0$. We
find that the coefficient $B(\vec{k})$ vanishes identically along this
line. Although the coefficient $A(\vec{k})$ is equal to $v_F k_y$ for
the zeroth order Hamiltonian, it is possible that for large values of
$\alpha_D$ it could pass through zero and change sign at one or more
values of $k_y$ other than $k_y =0$. If this occurs, and if
$C(\vec{k})$ were also zero, then there would be zero-energy states at
these values of $k_y$. However we shall see that along the line $k_x
=0$, the coefficient $C(\vec{k})$ is different from zero in third
order perturbation theory. Although $C(\vec{k})$ could have zeroes
along the $k_x =0$ axis for sufficiently large values of $\alpha_D$,
there is no symmetry reason why these should occur at the points where
$A(\vec{k})$ vanishes. 

Similarly we find that along the $k_y=0$ axis $B(\vec{k}) =
C(\vec{k})=0$ to all orders in perturbation theory, but that
$A(\vec{k})$ is generally non zero there. Isolated energy zeroes are 
therefore allowed by symmetry along the $k_y=0$ axis and will be
found if $A(\vec{k})$ vanishes at any point on this symmetry line. 

For other points in the Brillouin zone, neither $A$ nor $B$ nor $C$
vanish by symmetry, and there is no special relation between them. By
varying $k_x$ and $k_y$, one might find some isolated points where $A$
and $B$ vanish simultaneously; however there is no reason why $C$
should also vanish at such a point. Thus for a generic fixed value of
$\alpha_D$, there should be no further zero-energy points in the Brillouin 
zone, other than along the $k_y=0$ axis, where we find at least one state of 
zero energy at the $\Gamma$ point. By varying
$\alpha_D$, however, it is possible that one could find special values
where there are additional isolated zero-energy points.

To summarize the results of the perturbative analysis, we find that
there is always an energy zero at the $\Gamma$ point, for a Bravais
vortex lattice. In the case of a vortex lattice with a rectangular
unit cell, rotated by $45^{\circ}$ from the quasiparticle anisotropy
axis, there can be, for large enough anisotropy $\alpha_D$,
additional zero-energy states along the $k_y=0$ axis. Of course, this
result holds for any vortex lattice whose magnetic unit cell can
be chosen as a rectangular unit cell properly oriented, in particular  
the triangular vortex lattice. At any other point of the magnetic
Brillouin zone there will generally be no further zeroes in the energy
spectrum, although there could be very low energy states. Also, at
isolated values of the anisotropy ratio $\alpha_D$
there could be energy zeroes at non-symmetry points in the
Brillouin zone, but never lines of zero-energy states.

We now show explicitly that for $k_x =0$ the coefficient $B(\vec{k})$
is zero to all orders in perturbation theory, while $C(\vec{k})$
is nonzero at third order. We first consider the second order
effective Hamiltonian 
\begin{equation}
{\cal H}_{\mbox{\scr eff}}^{(2)}(E) = {\cal P}_{k_y}\left({\cal H}_0 + 
{\cal H}_1 \frac{1-{\cal P}_{k_y}}{E-{\cal H}_0} {\cal H}_1 \right)
{\cal P}_{k_y},
\label{HeffK}
\end{equation} 
where ${\cal P}_{k_y}$ is the projection operator onto the
space spanned by
$\{|E^{(0)}_{(0,k_y),+}\rangle,|E^{(0)}_{(0,k_y),-}\rangle\}$. 
Let us define $E^{(0)}_{k_y,\vecscr{Q},i} = E^{(0)}_{(Q_x,Q_y+k_y),i}$,
$\alpha_{k_y,\vecscr{Q},i} = \alpha_{(Q_x, Q_y+ k_y),i}$ and
$\beta_{k_y,\vecscr{Q},i} = \beta_{(Q_x, Q_y+ k_y),i}$. 
Analogously to (\ref{pert++}-\ref{pert--}), we can calculate the
matrix elements of this 2 by 2 matrix
\end{multicols}
\begin{eqnarray}
\langle E^{(0)}_{k_y,\vecscr{0},+}|{\cal H}_{\mbox{\scr
eff}}^{(2)}(E)| E^{(0)}_{k_y,\vecscr{0},+}\rangle &=&
E^{(0)}_{(0,k_y),+}   + \sum_{\vecscr{Q}\neq 0, i = \pm}
\frac{1}{E - E^{(0)}_{k_y,\vecscr{Q},i}} \left|
\alpha_{k_y,\vecscr{Q},i} 
{\cal V}^{(0)}_{\vecscr{Q}} + \alpha_{k_y,\vecscr{Q},i}
{\cal V}^{(3)}_{\vecscr{Q}}+\beta_{k_y,\vecscr{Q},i}
{\cal V}^{(1)}_{\vecscr{Q}}\right|^2 
\label{pertE++} \\
\langle E^{(0)}_{k_y,\vecscr{0},+}|{\cal H}_{\mbox{\scr eff}}^{(2)}(E)|
E^{(0)}_{k_y,\vecscr{0},-}\rangle &=& 
\sum_{\vecscr{Q}\neq 0, i = \pm}
\frac{1}{E-E^{(0)}_{k_y,\vecscr{Q},i}} \left( \alpha_{k_y,\vecscr{Q},i}
{\cal V}^{(0)}_{\vecscr{Q}} + \alpha_{k_y,\vecscr{Q},i}
{\cal V}^{(3)}_{\vecscr{Q}}+\beta_{k_y,\vecscr{Q},i}
{\cal V}^{(1)}_{\vecscr{Q}}\right) \nonumber \\
&& \times \left(\alpha_{k_y,\vecscr{Q},i} {\cal V}^{(1)}_{\vecscr{Q}}-
\beta_{k_y,\vecscr{Q},i}
{\cal V}^{(0)}_{\vecscr{Q}} - \beta_{k_y,\vecscr{Q},i}
{\cal V}^{(3)}_{\vecscr{Q}}
\right) \label{pertE+-} \\
\langle E^{(0)}_{k_y,\vecscr{0},-}|{\cal H}_{\mbox{\scr
eff}}^{(2)}(E)|E^{(0)}_{k_y,\vecscr{0},-}\rangle &=&
E^{(0)}_{(0,k_y),-}  + \sum_{\vecscr{Q}\neq 0, i = \pm}
\frac{1}{E-E^{(0)}_{k_y,\vecscr{Q},i}} \left|
\beta_{k_y,\vecscr{Q},i} 
{\cal V}^{(0)}_{\vecscr{Q}} - \beta_{k_y,\vecscr{Q},i}
{\cal V}^{(3)}_{\vecscr{Q}}+\alpha_{k_y,\vecscr{Q},i}
{\cal V}^{(1)}_{\vecscr{Q}}\right|^2. 
\label{pertE--} 
\end{eqnarray} 
\begin{multicols}{2} 
Looking at the off-diagonal terms (\ref{pertE+-}), we see that the
imaginary part of these matrix elements vanishes if the vortex lattice
is a Bravais lattice because, as we noted earlier,
${\cal V}^{(0)}_{\vecscr{Q}}$ vanishes when ${\cal V}^{(1)}_{\vecscr{Q}}$ and
${\cal V}^{(3)}_{\vecscr{Q}}$ do not and vice versa. Thus to second
order, the coefficient of $\sigma_2$ is zero. This property
holds for arbitrary $\vec{k}$, not just
on the $k_y-$axis, as $\alpha_{\vecscr{k},\vecscr{Q},i}$ and
$\beta_{\vecscr{k},\vecscr{Q},i}$ are real numbers for every $\vec{k}$
and even though the off-diagonal matrix element away from the $k_x =0$
axis has a more complicated structure than in (\ref{pertE+-}), its
imaginary part will still be a sum of polynomials in
$\alpha_{\vecscr{k},\vecscr{Q},i}$ and
$\beta_{\vecscr{k},\vecscr{Q},i}$ times ${\cal
V}^{(0)}_{\vecscr{Q}}{\cal V}^{(1,3)}_{\vecscr{Q}}$, which vanish
identically. 

Going back to the $k_x =0$ case, we can identify one further symmetry
$\alpha_{k_y,(-Q_x,Q_y),i} = \alpha_{k_y,(Q_x,Q_y),i}$ and
$\beta_{k_y,(-Q_x,Q_y),i} = - \beta_{k_y,(Q_x,Q_y),i}$ which makes the
$\sigma_1$ term in the effective Hamiltonian vanish $B(0,k_y)=0$.

If we look at
the third order off-diagonal matrix element of the effective
Hamiltonian, analogously to (\ref{third}), we have, at $E=0$: 
\end{multicols}
\begin{eqnarray}
&& \langle E^{(0)}_{k_y,\vecscr{0},+}|{\cal H}_{\mbox{\scr eff}}^{(3)}|
E^{(0)}_{k_y,\vecscr{0},-}\rangle  =  \nonumber \\
&& \sum_{\vecscr{Q}_1,\vecscr{Q}_2\neq 0, i_1,i_2 = \pm}
\frac{\left(\alpha_{k_y,\vecscr{Q}_1,i_1}
{\cal V}^{(0)}_{\vecscr{Q}_1} + \alpha_{k_y,\vecscr{Q}_1,i_1}
{\cal V}^{(3)}_{\vecscr{Q}_1}+\beta_{k_y,\vecscr{Q}_1,i_1}
{\cal V}^{(1)}_{\vecscr{Q}_1}\right)\left(-\beta_{k_y,\vecscr{Q}_2,i_2}
{\cal V}^{(0)}_{\vecscr{Q}_2} - \beta_{k_y,\vecscr{Q}_2,i_2}
{\cal V}^{(3)}_{\vecscr{Q}_2}+ \alpha_{k_y,\vecscr{Q}_2,i_2} {\cal
V}^{(1)}_{\vecscr{Q}_2 } \right)}{E^{(0)}_{k_y,\vecscr{Q}_1,i_1} E^{(0)}_{k_y,\vecscr{Q}_2,i_2}} \nonumber \\
&& ~~~~~ \times \left[{\cal V}^{(0)}_{\vecscr{Q}_2 -
\vecscr{Q}_1}(\alpha_{k_y,\vecscr{Q}_1,i_1} \alpha_{k_y,\vecscr{Q}_2,i_2} +
\beta_{k_y,\vecscr{Q}_1,i_1} \beta_{k_y,\vecscr{Q}_2,i_2}) 
\right. + {\cal V}^{(3)}_{\vecscr{Q}_2 - \vecscr{Q}_1} (\alpha_{k_y,\vecscr{Q}_1,i_1}
\alpha_{k_y,\vecscr{Q}_2,i_2} - \beta_{k_y,\vecscr{Q}_1,i_1}
\beta_{k_y,\vecscr{Q}_2,i_2}) \nonumber \\
&& ~~~~~~~~~~~~~~~~~~~~~~~~~~~~~~~~~~~~~~~~~~~~~~~~~~~~~~~~~~~~~~~~~ 
+ \left. {\cal V}^{(1)}_{\vecscr{Q}_2 -
\vecscr{Q}_1} (\alpha_{k_y,\vecscr{Q}_1,i_1} \beta_{k_y,\vecscr{Q}_2,i_2} +
\beta_{k_y,\vecscr{Q}_1,i_1} \alpha_{k_y,\vecscr{Q}_2,i_2}) \right] 
\label{thirdE}
\end{eqnarray}
\begin{multicols}{2}
Let us start by analyzing the coefficient of $\sigma_1$.
Because of the Bravais lattice symmetry, the matrix element
(\ref{thirdE}) is pure imaginary, and so can only contribute to
$\sigma_2$. This is true for every correction to the off-diagonal
matrix element of the effective Hamiltonian coming from odd orders of 
perturbation theory: there will always be an odd power of ${\cal
V}^{(0)}$ Fourier coefficients in every term in the sum over
intermediate states. The fourth order (and every other even order)
correction could have a real part coming from terms with $\vec{Q}_1$ 
odd (meaning that if $\vec{Q} = (2 \pi/d)
(m \hat{x}+ n \hat{y})$, then $m+n$ is odd) and every other
reciprocal lattice vector alternating between even and odd.
Considering pairs of terms of this kind with opposite $Q_x$ for every
reciprocal lattice vector of the intermediate states, it is easy to
see that they will have opposite signs using the symmetry properties
of the potential and of the unperturbed wave functions mentioned
above. In particular, this implies that the $\sigma_1$ term will keep
vanishing along the $k_x =0$ axis to every order in perturbation
theory. 

Finally, we turn to the $\sigma_2$ term. This time, the Bravais lattice
symmetry will ensure that corrections coming from even orders in
perturbation theory are real numbers and so the only contributions to
$\sigma_2$ can come from odd orders in perturbation theory. The third
order matrix element 
(\ref{thirdE}) is the lowest non-vanishing one and it can be
approximately evaluated numerically as a function of $k_y$ and the anisotropy
ratio $\alpha_D$. We find that it is generically non-zero when $k_y \neq 0$.

The perturbative analysis for the $k_y=0$ axis proceeds along the same
lines. In this case, besides the Bravais nature of the vortex lattice,
the symmetry responsible for the vanishing of the coefficients
$B(\vec{k})$ and $C(\vec{k})$ is $\alpha_{k_x,(Q_x,-Q_y),\pm} = -
\alpha_{k_x,(Q_x,Q_y),\mp}$ and $\beta_{k_x,(Q_x,-Q_y),\pm} = 
\beta_{k_x,(Q_x,Q_y),\mp}$. The matrix elements of the effective
Hamiltonian can be explicitly evaluated taking
$\alpha_{k_x,\vecscr{0},\pm}=\pm\frac{1}{\sqrt{2}}$ and
$\beta_{k_x,\vecscr{0},\pm}=\frac{1}{\sqrt{2}}$.      

To summarize, in this section we showed that for a Bravais lattice of vortices
with a rectangular unit cell and for generic values of the
anisotropy ratio $\alpha_D$, zero-energy states can only be found along the 
$k_y=0$ axis (one Dirac point is always found at the $\Gamma$ point). Anywhere
else in the magnetic Brillouin zone we do not 
expect to find further Dirac points, even though the gaps separating the 
particlelike and holelike bands could get very small. Also, for special 
values of $\alpha_D$ it is not ruled out that there could be isolated energy 
zeroes anywhere in the Brillouin zone.

\section{Vortex lattice with a basis}
\label{sect:epsilon}
\subsection{Two vortices per unit cell}
As we noted earlier, the key ingredient to find a gapless spectrum at the
center of the Brillouin zone in perturbation theory was the Bravais
nature of the vortex lattice. We can 
explore this connection further relaxing the constraint on the
position of the vortices. The unit cell had to be doubled in order to
enclose one quantum of magnetic flux $\Phi_0 = hc/|e|$; furthermore we
divided the vortices into two sublattices $A$ and $B$ but kept them
evenly spaced. We can leave the geometry of the two sublattices
unchanged but displace them with respect to each other thus changing
our lattice into a non-Bravais lattice with two vortices per unit
cell. Let us assume that the two sublattices $A$ 
and $B$ are still square lattices with spacing $d$, but let us consider
what happens if we let the distance between nearest $A$ and $B$
vortices be different from $(\sqrt{2}/2) d$. For concreteness let us take
\begin{equation}
\vec{R}_0 = \left(\frac{1-\epsilon_x}{4}
\hat{x}+\frac{1-\epsilon_y}{4} \hat{y}\right) d,
\label{p}
\end{equation}
as defined in Fig.~\ref{figAB}. Then, 
$\vec{Q} \cdot \vec{R}_0 = (\pi/ 2) [(1-\epsilon_x) m + (1-\epsilon_y)
n]$ and 
\begin{equation}
{\cal V}^{(0)}_{-\vecscr{Q}} \pm {\cal V}^{(3)}_{-\vecscr{Q}} = ({\cal
V}^{(0)}_{\vecscr{Q}} \pm {\cal V}^{(3)}_{\vecscr{Q}})^*.
\end{equation}
The diagonal matrix elements of the effective Hamiltonian
at the $\Gamma$ point (\ref{pert++}) and (\ref{pert--}), will keep
vanishing because of the 
symmetries outlined above. On the other hand the off-diagonal terms do
not vanish anymore and are purely imaginary, the second order
effective Hamiltonian is thus proportional to $\sigma_2$. The series
$(\ref{pert+-})$ can be written as (keeping only the non-vanishing
imaginary part)
\end{multicols}
\begin{eqnarray}
&& \langle E^{(0)}_{0,+} | {\cal H}_{\mbox{\scr eff}}^{(2)} |
E^{(0)}_{0,-} \rangle = 
\sum_{\vecscr{Q}\neq 0, i = \pm}
\frac{{\cal V}^{(0)}_{\vecscr{Q}}}{E^{(0)}_{\vecscr{Q},i}} \left[ 2
\alpha_{\vecscr{Q},i} \beta_{\vecscr{Q},i} {\cal V}^{(3)}_{\vecscr{Q}}
+(\beta_{\vecscr{Q},i}^2 - \alpha_{\vecscr{Q},i}^2)
{\cal V}^{(1)}_{\vecscr{Q}}\right] \nonumber \\
&& = i \left(\frac{\hbar v_F}{d}\right)^2 \sum_{(m,n)\neq (0,0), i = \pm} 
\frac{(-1)^{m+n+1}}{2
E^{(0)}_{\vecscr{Q},i}} \frac{m}{(m^2+n^2)^2} \sin \pi (\epsilon_x m +
\epsilon_y n) \left[2 \alpha_{\vecscr{Q},i} \beta_{\vecscr{Q},i} m -
\frac{1}{\alpha_D}(\beta_{\vecscr{Q},i}^2 - \alpha_{\vecscr{Q},i}^2)
n \right].
\label{+-matrixel}
\end{eqnarray}
\begin{multicols}{2}
For small $\vec{\epsilon}$ we can expand the previous expression to
first order in $\epsilon_x$ and $\epsilon_y$ and using the symmetry
\begin{equation}
\begin{array}{c}
\alpha_{(-m,n),i} = \alpha_{(m,n),i} \\  
\beta_{(-m,n),i} = - \beta_{(m,n),i}
\end{array}
\end{equation}
it is easy to show that the series does not depend on
$\epsilon_y$. This result implies that, contrary to what we found in
the Bravais lattice case, if the unit cell of the vortex lattice has a
basis composed of two vortices with $\epsilon_x \neq 0$, the
quasiparticle spectrum becomes gapped. In second order perturbation theory, the
lowest eigenvalue at the center of the Brillouin zone depends linearly on
$\epsilon_x$ for small distortions
\begin{equation}
E_{\mbox{\scr gap}}^{(\vecscr{\epsilon})} = \frac{\hbar v_F}{d}
g(\alpha_D) |\epsilon_x|
\end{equation}
where the function $g(\alpha_D)$ is defined as
\begin{eqnarray}
\nonumber
 g(\alpha_D) &=&  \pi \frac{\hbar v_F}{d} \sum_{
(m > 0,n), i = \pm} \frac{(-1)^{m+n+1}}{
E^{(0)}_{\vecscr{Q},i}} \frac{m^2}{(m^2+n^2)^2} \\ &\times&
\left[2 \alpha_{\vecscr{Q},i} \beta_{\vecscr{Q},i} m -
\frac{1}{\alpha_D}(\beta_{\vecscr{Q},i}^2 - \alpha_{\vecscr{Q},i}^2)
n \right].
\end{eqnarray}

The asymmetry between $\epsilon_x$ and $\epsilon_y$ (present even in
the isotropic $\alpha_D =1$ case) is due to the term proportional to
the identity matrix in the Hamiltonian (\ref{Hdirac}), in fact the
terms in the series in (\ref{+-matrixel}) are proportional to ${\cal
V}_{\vecscr{Q}}^{(0)}$. Linearizing the Bogoliubov--de Gennes equation
around the Dirac points $\vec{p} = (p_F,0)$ or $\vec{p} = (-p_F,0)$
we would exchange the role of $x$ and $y$ so that for every distortion
$\vec{\epsilon}$ the total density of states, defined as the sum of
the density of states from the four Dirac points, exhibits the
fourfold symmetry of the vortex lattice explicitly.       

In order to compare perturbation theory to the exact numerical
diagonalization of the linearized Bogoliubov--de Gennes equation
(\ref{Hdirac}), 
let us consider the case $\epsilon = \epsilon_x = \epsilon_y$. The
matrix element (\ref{+-matrixel})  can be evaluated numerically and
for the case $\alpha_D =1$ (isotropic Dirac cone) and $\epsilon=0.1$ we find
\begin{equation}
{\cal H}_{\mbox{\scr eff}}^{(2)} = - 0.055 \frac{\hbar v_F}{d} \sigma_2
\end{equation}
while for $\epsilon=0.2$ we have
\begin{equation}
{\cal H}_{\mbox{\scr eff}}^{(2)} = - 0.11 \frac{\hbar v_F}{d} \sigma_2.
\label{eps0.2}
\end{equation}
The linear dependence of the gap on $\epsilon$ is evident.

We have run exact numerical diagonalization of the Hamiltonian
(\ref{Hdirac}) and have found for the lowest energy eigenvalue in the
$\epsilon =0.1$ case $0.0765 \hbar v_F/d$ while in the $\epsilon=0.2$
case $0.142 \hbar v_F/d$ which are in good agreement with the second
order perturbation theory results. These gaps are larger than the gap
induced by the Wilson term in the band structure in a Bravais vortex
lattice (where perturbation theory predicted gapless behavior) and
scale linearly with $\epsilon$ once the $\epsilon =0$ residual Wilson
gap is subtracted (for $\alpha_D=1$ this is approximately $0.010 \hbar
v_F/d$). 

\begin{figure}[t]
\noindent
\epsfxsize=8.5cm
\epsffile{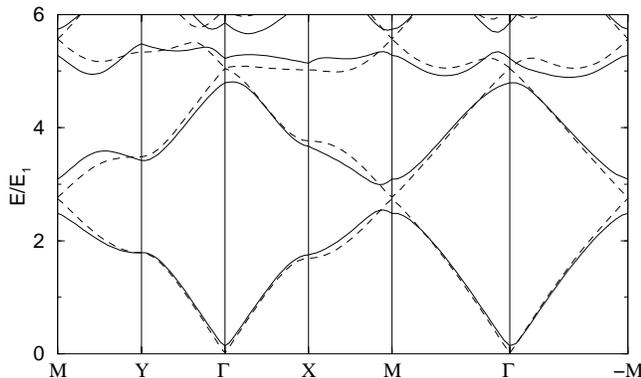}
\caption{Band structure for a non-Bravais lattice with two vortices
per unit cell in the $\epsilon_x = \epsilon_y = 0.2$, $\alpha_D=
v_F / v_\Delta = 1$ case (solid curve). 
The dashed curve is the spectrum for a simple square lattice
($\epsilon_x = \epsilon_y =0$). The gaps at the $\Gamma$ and M point
are real: they are much larger than the Wilson term contribution for anisotropy
$\alpha_D=1$, as can be seen comparing the dashed and solid curves. 
Notice that the symmetries $k_x \to -k_x$ and $k_y \to -k_y$ 
are broken in this case (the bands along the M$\Gamma$ and 
-M$\Gamma$ lines are different). Since particle-hole symmetry is still
preserved at each point in the Brillouin zone, only the positive energy 
bands are plotted.}
\label{figp}
\end{figure}

We also calculated the full band structure for the isotropic $\alpha_D
= 1$, $\epsilon=0.2$ case and the results are shown in
Fig.~\ref{figp}. Notice the gaps that open at the $\Gamma$ and M
points, while the rest of the band structure is qualitatively
unchanged, as one would expect from conventional perturbation
theory. Contrary to the square lattice case, the band structure is not
symmetric under the exchange of $k_x$ with $-k_x$ or $k_y$ with $-k_y$
as can be seen from the energy bands in Fig.~\ref{figp}.

The density of states in the Bravais vortex lattice will be discussed
in detail in the next section but it is clear that the gapless
behavior at the $\Gamma$ point 
implies the existence of a low-energy window where the
density of states is linear and vanishes at zero energy, just like in
a homogeneous $d$-wave superconductor. If the
vortex lattice is distorted in the way discussed above, we
find a gap in the quasiparticle excitation spectrum which depends
on the magnitude and orientation of the distortion. The density of
states will then vanish at zero energy in the general case of a vortex
lattice with two vortices per unit cell.

\subsection{Four vortices per unit cell}
In previous sections of the paper we have discussed the role played by
particle-hole symmetry in determining some of the key features of the
spectrum. In particular, we noticed that the vanishing of the density
of states at zero energy, with or without a gap opening at the center
of the Brillouin zone, is deeply related to this symmetry. To
explore this connection further, it is of interest to find more complicated
vortex lattice structures that break particle-hole symmetry and
see the effect on the spectrum. 

\begin{figure}[t]
\noindent
\epsfxsize=8.5cm
\epsffile{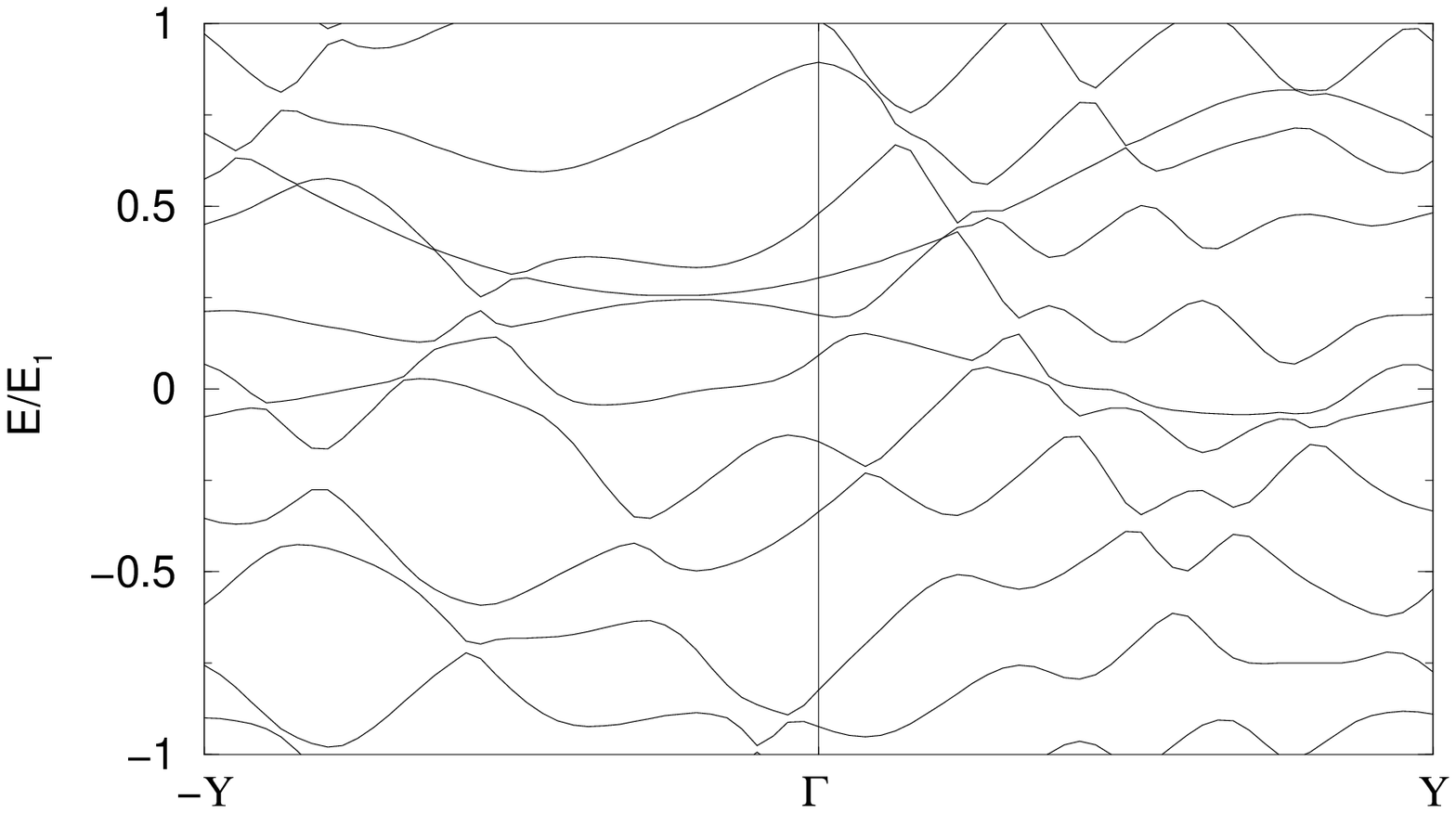}
\\*
\noindent
\epsfxsize=8.5cm
\epsffile{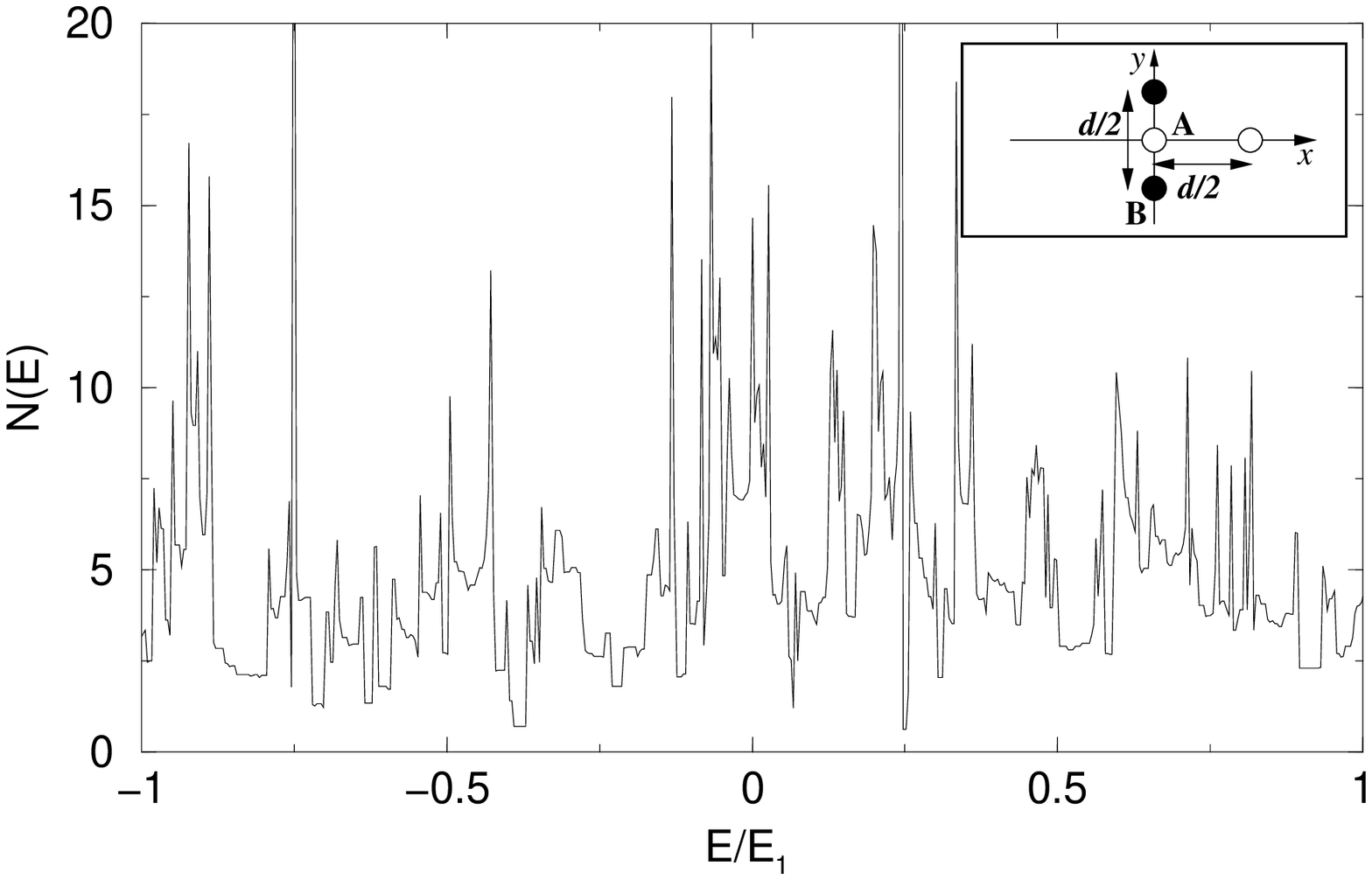}
\caption{Band structure and density of states for a non-Bravais lattice with 
four vortices
per unit cell with anisotropy ratio $\alpha_D
= v_F / v_\Delta = 5$. The position of the vortices in the unit cell
is displayed in the inset, notice that there is no center of inversion
symmetry and thus particle hole symmetry is absent.}
\label{figfournosym}
\end{figure}

Particle-hole symmetry requires a center of inversion in the
unit cell to exist. We can break this symmetry considering a unit cell
with a basis consisting of four vortices, so that its area is 
$2 d^2$. We choose a rectangular unit cell with sides $-d < x < d$,
$-d/2<y< d/2$. As an example, we can study the quasiparticle spectrum when 
the A-vortices are
located at $(0,0)$ and $(d/2,0)$ and the B-vortices are located at 
$(0,\pm d/4)$, as shown in the inset in Fig.~\ref{figfournosym}. 
In this case bands can go through the zero-energy
axis away from crossings or near-crossings and thus the discussion in
Section~\ref{sect:pertgen} doesn't hold anymore and lines of zeroes
can be found. The band structure in Fig.~\ref{figfournosym} for
anisotropy $\alpha_D = 5$ shows such lines of zero-energy states. 
Instead of going to zero, the
density of states stays finite all the way down to zero energy, in
close analogy to the prediction of the semiclassical theory
\cite{volovik93}. It may also be seen that in this example, there is
no particle-hole symmetry for the overall density of states for the
exhibited band structure. We must recall, however, that the exhibited
states are derived from only one of the four Dirac points, $\vec{p}=(0,p_F)$,
of the zero-field Fermi surface. If we include the contribution from
the opposite point $\vec{p}=(0,-p_F)$ the overall particle-hole symmetry will
be recovered.

\begin{figure}[t]
\noindent
\epsfxsize=8.5cm
\epsffile{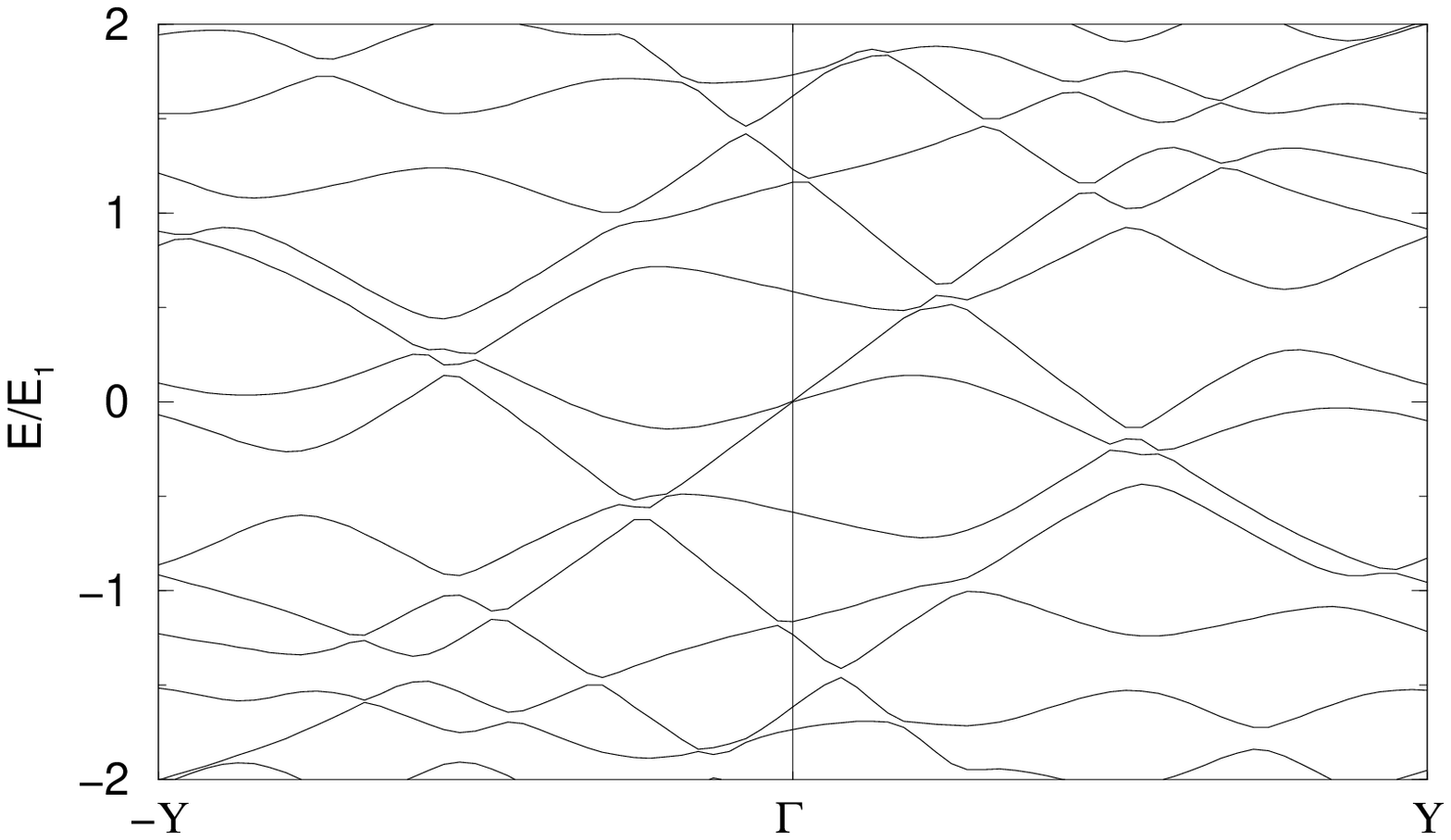}
\\*
\noindent
\epsfxsize=8.5cm
\epsffile{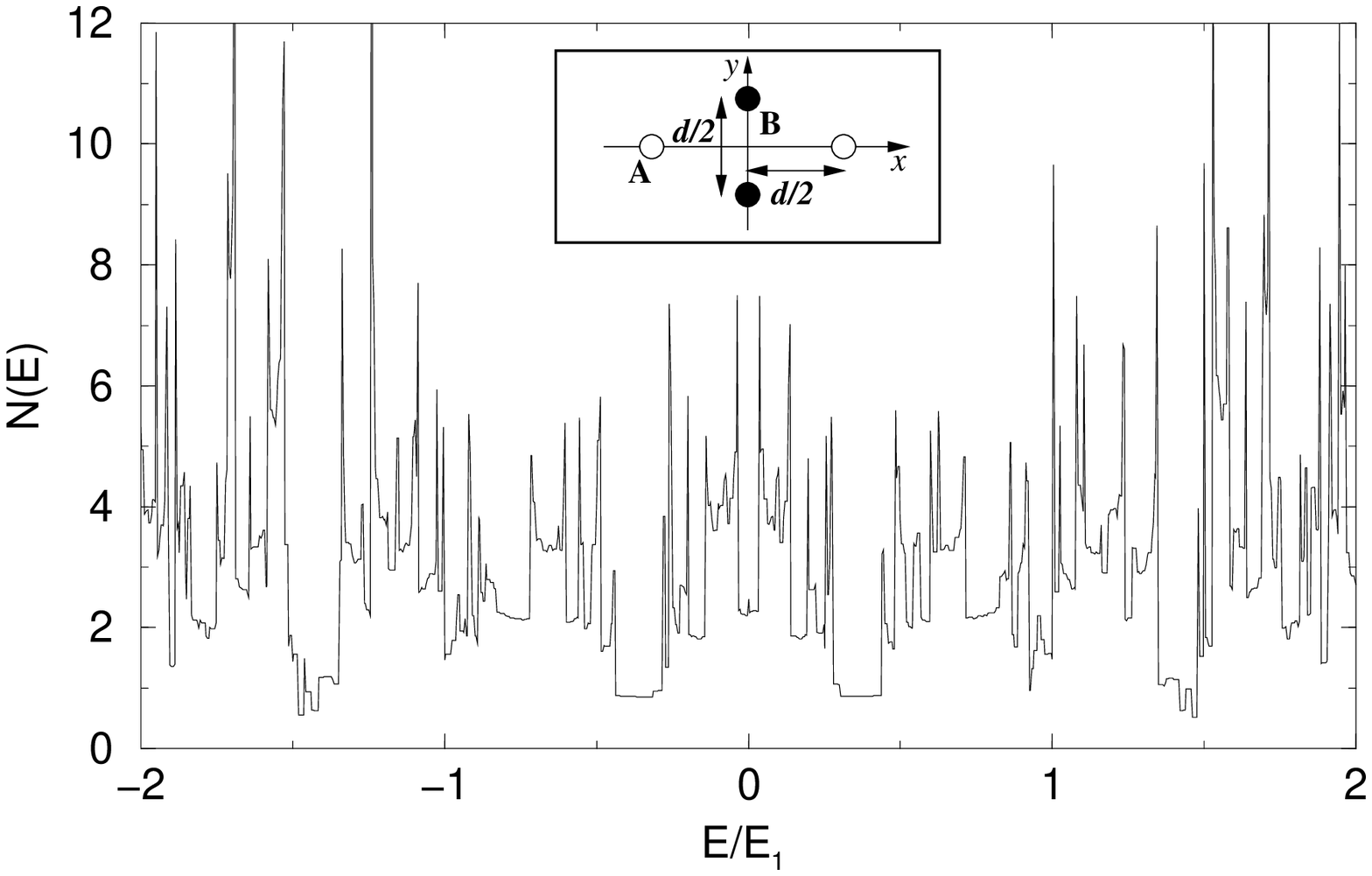}
\caption{Band structure and density of states for a non-Bravais lattice 
with four vortices per unit cell with anisotropy ratio $\alpha_D
= v_F / v_\Delta = 5$. The position of the vortices in the unit cell
is displayed in the inset, notice that there is a center of inversion
symmetry with respect to which $\vec{v}^{A,B}(-\vec{r}) = - 
\vec{v}^{A,B}(\vec{r})$.  
Particle-hole symmetry is preserved, although not independently at every 
point in the Brillouin zone: states at $\vec{k}$ and $-\vec{k}$ have to be 
exchanged.}
\label{figfoursym}
\end{figure}

With four vortices per unit cell it is also possible to find superfluid
velocity distributions that preserve particle-hole symmetry for the total 
density of states arising from a single Dirac point, but
not at each point of the Brillouin zone separately as was the case
for two vortices per unit cell. For example, the
configuration depicted in Fig.~\ref{figfoursym}, where the 
A-vortices are located at $(\pm d/2,0)$ while the B-vortices are at the same 
position as before, shows such a distribution. 
Here, if we consider the transformation $\vec{r}
\to - \vec{r}$, the superfluid velocities $\vec{v}^A(\vec{r})$ and
$\vec{v}^B(\vec{r})$ are not exchanged as in equation (\ref{vtransf}),
rather they transform like $\vec{v}^{A,B}(-\vec{r}) =
-\vec{v}^{A,B}(\vec{r})$. The Hamiltonian (\ref{HBloch}) goes into
minus itself if we take $\vec{r} \to -\vec{r}$ and simultaneously
exchange $\vec{k}$ with $-\vec{k}$. We then find $E_+(\vec{k}) = -
E_-(-\vec{k})$, but no particle-hole symmetry at a fixed $\vec{k}$. 
The density of states is an even function of the quasiparticle energy,
as can be observed in Fig.~\ref{figfoursym}. Lines of energy-zeroes are still 
allowed in this case, so the density of states may be finite at $E=0$.

\section{Density of states: comparison with the semiclassical theory}
\label{sect:dos}
The density of states is computed using a linear
interpolation for the band structure in between the sampled
$k-$points and is normalized as follows:
\begin{equation}
N(E) = 2 \sum_n \int \frac{d^2k}{(2 \pi)^2} \delta(E-E_n(\vec{k}))
\end{equation} 
where the factor of 2 comes from spin degeneracy and $n$ is a band
index. As we noted in the
previous section, this is the contribution to the total density of
states coming from one of the four nodes of the zero-field
quasiparticle spectrum. For a simple square vortex lattice in the 
orientation we are considering, the total density of states can be
obtained simply by multiplying this result by four. In more
complicated vortex lattices, a
separate calculation of the density of states at one of the nodes rotated by
$90^\circ$ with respect to $\vec{p} = (0,p_F)$ is generally necessary.

If the vortex lattice is a Bravais lattice, the quasiparticle spectra
are gapless regardless of the anisotropy ratio $\alpha_D$ and the
density of states at very low-energy is
linear in energy although the slope is renormalized by the
potentials (and the renormalization factor depends on $\alpha_D$). 
The results of the numerical diagonalization are shown in
Figs.~\ref{figdelta1}-\ref{figdelta4} for $\alpha_D = 1, 2, 4$
respectively. 
The sharp peaks in the density of states are logarithmic
van Hove singularities: for topological reasons every band in two
dimensions has at least two saddle points which  
contribute logarithmically divergent peaks to the density of
states. The van Hove singularities show up as finite-height peaks in
the numerical evaluation of the density of states because of the
linear interpolation scheme used for the band structure. 
Averaging these peaks 
out, however, one can see that at high energy the density of states reproduces
the behavior expected for the quasiparticles in the absence of a
magnetic field $N(E)= |E|/(\pi \hbar^2 v_F v_\Delta)$ as shown in 
Fig.~\ref{figdelta4} for the $\alpha_D=4$ case.

We want to compare our results to the semiclassical picture studied 
primarily by Volovik \cite{volovik93,kopnin96}. This approach takes 
into account the superfluid
velocity $\vec{v}_s$ distribution through the Doppler shift of the 
quasiparticle energy
\begin{equation}
E(\vec{k},\vec{r}) = \pm \sqrt{\xi^2_{\vecscr{k}}+\Delta_d(\vec{k})^2}
+ \hbar \vec{k}\cdot \vec{v}_s(\vec{r}),
\label{scenergy}
\end{equation}
where $\xi_{\vecscr{k}}$ is the kinetic energy measured with respect
to the Fermi surface. Within this framework Kopnin and Volovik introduced
two crossover energy scales \cite{kopnin96,volovik97,volovik97b} $E_1
\approx \hbar v_F/d$ and $E_2^{KV} \approx \hbar v_\Delta/d$. The first
energy scale $E_1 \approx \hbar v_F/d$ marks the boundary between the
temperature dominated regime and the
superflow dominated regime in the thermodynamic functions. Physically,
this crossover corresponds to the WKB eigenfunctions becoming
extended on a scale comparable to the intervortex distance, at least
in one direction.  For $E > E_1$, the
states are unaffected by the magnetic field and the density of states
is linear with the same slope one would find in the bulk without a
vortex lattice, as we discussed in the previous paragraph. 
For $E< E_1$, the semiclassical density of states is
essentially independent of energy \cite{volovik93} and for our order parameter
distribution on a square lattice we calculate it to be
\begin{equation}
N(0) = \frac{1}{4 \hbar v_\Delta d}.
\label{n(0)}
\end{equation}
Volovik finds essentially the same result \cite{volovik93}, with an 
undetermined numerical prefactor which depends on the geometry of the
vortex lattice. Won and Maki \cite{won96}, using a somewhat different
model, calculate this geometric prefactor for a general vortex lattice
structure (although only Bravais lattices are considered) and find a
result of the same order of magnitude of (\ref{n(0)}), for a square lattice.

The second crossover marks the boundary where a full quantum
mechanical picture becomes important and the semiclassical analysis
breaks down. We will call this scale $E_2$. As we mentioned above,
Kopnin and Volovik \cite{kopnin96,volovik97,volovik97b} argue that
this scale is linear in $1/\alpha_D$, and is given by $E_2^{KV} \approx \hbar
v_\Delta/d$ and, for energies in the range $E_2^{KV}<E<E_1$, they predict a
constant density of states. In terms of band structure this means that
we need to find a direction in $\vec{k}$-space in which the bands are
flat on a scale of $E_2$. If we assume that the perturbation induced
by the magnetic field is weak, and thus that the band structure is only
weakly renormalized, we find that for large enough anisotropy
$\alpha_D \gg 1$ several 
bands will start overlapping for energies $E<E_1$ and will have a
small dispersion in the $k_x$ direction of the order of $\hbar
v_\Delta/d$. The density of states for $E<E_1$ would then be
essentially constant down to energies of the order of $\hbar v_\Delta/d$,
which corresponds exactly to $E_2^{KV}$. For energies $E< E^{KV}_2$ we
would have just a single band and the density of states would drop
linearly to zero. The key assumptions that enter in
this argument (weak perturbing potential) are essentially the absence
of vanishing energies, or nearly vanishing energies, at any point 
in the Brillouin zone other
than the $\Gamma$ point and a small renormalization of the slope of
the lowest energy bands at the $\Gamma$ point in any direction of the
Brillouin zone. These assumptions are satisfied for rather small
anisotropy ratios $\alpha_D \ll 10$ but seem to fail for higher
values of $\alpha_D$, as we will show.

In Fig.~\ref{figdelta1}-\ref{figdelta4} we can see how the
semiclassical description starts developing. For $\alpha_D=1$ there is no
resemblance to the semiclassical behavior for any energy, the bands are
very distinct at low energy and there is only one crossover from
a quantum mechanical region to a purely classical region ($E>E_1$) without any
hint of constant density of states. Changing the anisotropy to
$\alpha_D = 2$ or even better $\alpha_D= 4$ a hint of the semiclassical region
starts opening up and one can identify a trend towards a flat density
of states between roughly $E_2^{KV}$ and $E_1$. For energies much
lower than $E_2^{KV}$ the density of states is linear and goes to zero
at zero energy. Although there is only one band (even in the $\alpha_D
= 4$ case) below $E_1$, the semiclassical description seems to start
working remarkably well. From these plots we can already notice
discrepancies with the Kopnin and Volovik picture. Notice that, while
for the $\alpha_D =2$ case the ratio between the slopes in the $k_y$
and $k_x$ directions is $v_F^R /v_\Delta^R \approx 2.5 \sim \alpha_D$ (the
superscript $R$ indicates that these are the renormalized velocities
in the two above mentioned directions in $\vec{k}$-space), already in the
$\alpha_D =4$ case the same ratio is roughly 17 which is much larger than
$\alpha_D$. The scale at which the flat density of states should break
down, $E_2$, seems to be much smaller than the simple argument above
would predict.

\begin{figure}[t]
\noindent
\epsfxsize=8.5cm
\epsffile{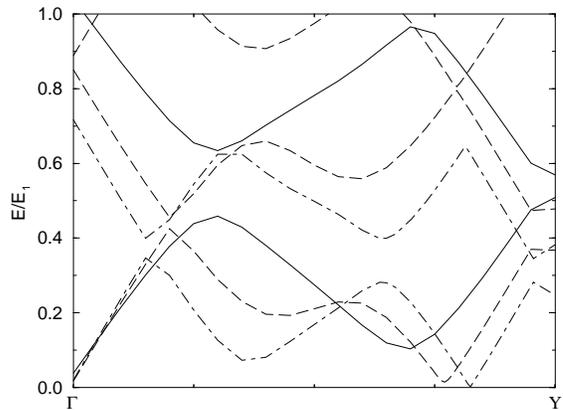}
\caption{Band structure for a square lattice with anisotropy ratio $\alpha_D
= v_F / v_\Delta =$ 8, 12, 15 (solid, dashed and dot-dashed lines
respectively) along the symmetry line
$\Gamma$Y. Notice the apparent vanishing of the energy spectrum at a
wavevector close to the Y point, for the large values of $\alpha_D$.
(Our numerical calculation cannot distinguish between a true zero and
a very small energy gap.) A second small energy gap is developing
about one third of the way from $\Gamma$ to Y. 
Only the positive energy bands 
are plotted for clarity, the negative energy ones can 
be obtained by particle-hole symmetry. In the $\alpha_D =15$ only the 
two lowest bands are plotted, while all the bands with energy $E<\hbar
v_F /d$ are 
plotted for $\alpha_D=$ 8, 12. The gaps at the $\Gamma$ point are fictitious 
and are due to the Wilson term in the Hamiltonian.}
\label{figalpha}
\end{figure}

Besides a large renormalization of the slope of the energy bands at
the $\Gamma$ point, which occurs even for relatively small
anisotropies, we find for large anisotropies that there are lines in
the Brillouin zone where the energy of the lowest band is very close
to vanishing. As we discussed in Section \ref{sect:pertgen}, we do not
expect to find points, other than the $\Gamma$ point and possibly
along the $k_y=0$ axis, where the energy
is exactly zero (except, conceivably, at isolated values of the
anisotropy ratio $\alpha_D$).
For anisotropies $\alpha_D > 8$ we cannot resolve numerically any
dispersion along the $k_x$ direction, so only the $\Gamma$Y line of
the band structure carries information. In Fig.~\ref{figalpha} we have
plotted a few of the lowest energy bands corresponding to values of
the anisotropy ratio $\alpha_D = $8, 12, 15. One can immediately see
one of these energy near-zeroes developing along the $\Gamma$Y axis. 
In general, the crossover scale $E_2$ will be set by the
larger of the energy gap at this point and the energy dispersion in
the $k_x$ direction. The density of states will be very close
to a constant for energies larger than $E_2$ and will drop towards zero
with decreasing energy for $E \ll E_2$. In this way the
high-anisotropy limit approaches the semiclassical prediction much faster
then linearly in $1/\alpha_D$. (The precise functional
dependence of $E_2$ on $\alpha_D$ will be discussed in a later publication.) 

For more complicated lattices, where there is not particle-hole
symmetry at each point in the Brillouin zone, there is no argument to
prevent zero crossings, and we do indeed find lines of zeroes for
large values of $\alpha_D$ (see Figs. \ref{figfournosym} and
\ref{figfoursym}). Thus the density of states is finite at
zero energy and the semiclassical results may apply down to zero energy.

\section{Conclusions}
\label{sect:concl}
In conclusion, we have studied the quasiparticle spectrum of a
$d$-wave superconductor in the mixed state. One important step in
solving the problem has been the transformation due to Franz and
Te\v{s}anovi\'{c} \cite{franz00} that maps the original
Bogoliubov--de Gennes equation into a Dirac Hamiltonian in an effective  
periodic vector and scalar potential corresponding to zero average 
magnetic field. We have found both numerically and in perturbation
theory that for a Bravais lattice of vortices the spectrum remains
gapless when a magnetic field is turned on. We have showed that for
vortex lattices which preserve the particle-hole symmetry of the
energy spectrum there
can only be other isolated Dirac points in the Brillouin zone and so they
cannot change qualitatively the very low-energy density of
states. Different conclusions are reached when more complicated vortex
lattice structure (for example with four vortices per unit cell) are
considered where lines of energy-zeroes can be found. In this case,
the density of states is finite at zero energy for large enough
anisotropy ratio $\alpha_D$. 
A non-Bravais vortex lattice with two-vortices per unit cell
can break the symmetry that keeps the spectrum gapless and open gaps
whose magnitude depends on both the magnitude and orientation of the
distortion observable both in the numerics and perturbation theory, with good
agreement between the two. Finally, the high-anisotropy limit has been
investigated and it's relation to the semiclassical analysis
explained. The crossover scale between the semiclassical and quantum
mechanical regime $E_2$ goes to zero for large values of the
anisotropy much faster than linearly in $1/\alpha_D$, at least for the
vortex lattice geometries considered here, and the density
of states quickly approaches a constant value for energies $E_2 < E <
E_1$.

\section{Acknowledgments}
We are grateful for helpful discussions with Noam Bernstein and Greg
Smith, and for support from NSF grant DMR 99-81283.

\vspace*{-10pt}


\end{multicols}

\end{document}